\documentclass{jfm}
\usepackage{amsmath}
\usepackage{graphicx}
\usepackage{multirow}
\usepackage{natbib}
\graphicspath{{./responseFigs/}}
\usepackage{color}

\newcommand{\Ca}{C_A}
\newcommand{\aoa}{\phi}
\newcommand{\wdot}{\dot{w}}
\newcommand{\Wdot}{\dot{W}}

\newcommand{\bu}{\boldsymbol{u}}
\newcommand{\bx}{\boldsymbol{x}}
\newcommand{\bF}{\boldsymbol{F}}
\newcommand{\bE}{\boldsymbol{e}}
\newcommand{\bn}{\boldsymbol{n}}
\newcommand{\perm}{a}
\newcommand{\rot}{b}

\newcommand{\defaoa}{\theta_\text{def}}
\newcommand{\vort}{\omega}
\newcommand{\vortf}{\omega_f}
\newcommand{\vortb}{\omega_b}
\newcommand{\vortbo}{\gamma}
\newcommand{\grad}{\boldsymbol{\nabla}}
\newcommand{\zhat}{\hat{\boldsymbol{e}_z}}

\newcommand{\xhat}{\hat{\boldsymbol{e}_x}}
\newcommand{\rhat}{\hat{\boldsymbol{e}_r}}
\newcommand{\thetahat}{\hat{\boldsymbol{e}_{\theta}}}
\newcommand{\diff}{\text{d}}
\newcommand{\zbar}{\bar{z}}
\newcommand{\zobar}{\bar{z}_0}
\newcommand{\zeroflow}{\left(\dfrac{\sqrt{1+4\tan^2\aoa} - 1}{2} \right)^{1/2}}
\newcommand{\etamax}{\eta_\text{max}}

\newcommand{\LBJ}{BJ}

\graphicspath{{Figures/}}
\newcommand{\etaval}{ 0.57}
\newcommand{\etavalmin}{ 0.39}
\newcommand{\etapercent}{ 57}
\newcommand{\etapercentmin}{ 39}

\title{Framework for systematic flow manipulation by wind and hydrokinetic energy turbine arrays}
\author{Shreyas Mandre\aff{1}\corresp{\email{shreyas\_mandre@brown.edu}} \and Niall M. Mangan\aff{1}\footnote{current address: Department of Engineering Sciences and Applied Mathematics, Northwestern University, Evanston, IL 60208}}
\affiliation{\aff{1} School of Engineering, Brown University, Providence RI 02912}

\begin{document}
\maketitle
\begin{abstract} 
Wind and hydrokinetic energy turbines are often constrained to locations where the available energy is limited by the operation of the turbines themselves.
In two-dimensions, we describe how an array can manipulate the steady flow, redirecting more fluid kinetic energy to itself.
Two computational examples of turbine arrays present solutions of the Navier-Stokes equations to illustrate the feasibility of flow manipulation, and motivate an idealized model.
Using inviscid fluid dynamics, we underscore the relation between bound vorticity and flow deflection, and between free vorticity and energy extraction.
To understand and design flow manipulations that increase the kinetic energy incident on the turbines, we consider an idealized deflector-turbine array constrained to a line segment, acting as an internal flow-boundary.
We impose profiles of bound and shed vorticity on this segment by parameterizing the flow deflection and the wake deficit, respectively, and analyze the resulting flow using inviscid fluid dynamics.
We find that the power extracted by the array is the product of two components: (i) the deflected kinetic energy incident on the array, and (ii) the array efficiency, both of which vary with deflection strength.
The array efficiency, or its ability to extract a fraction of the incident energy, decreases slightly with increasing deflection from about 57\% at weak deflection to 39\% at high deflection.
This decrease is outweighed by increasing incident kinetic energy with deflection, resulting in monotonically improved power extraction, thus highlighting the benefits from flow deflection.
\end{abstract}


\section{Introduction}
Practical considerations prevent the placement of tidal, riverine, and wind turbines at locations in the flow where they can extract maximum power.
For riverine and tidal power, the hotspots of fast flow occur at narrow constrictions in the channel, and anything less than a complete fence of turbines \citep{Garrett2005} across the flow constriction greatly reduces the potential of the site \citep{Vennell2010, Vennell2011, Cummins2013}. 
However, navigation and environmental considerations prohibit construction of such arrays \citep*{Garrett2007,Vennell2015}. 
Similarly, wind turbines cannot be constructed at locations of faster wind speed high up in the atmosphere due to structural constraints \citep*{Thresher2007}.
Instead, the most practical arrangement of wind turbines is an array of turbines mounted on the ground in rows -- in a plane parallel to the freestream -- where only turbines at the front of the array have access to direct flow and energy transport to turbines in the center of the array is limited \citep*[e.g., see][]{Porte-Agel2013}.
Any large riverine or tidal array in this configuration would suffer from the same limitations \citep*{Vennell2015,Jeffcoate2016}.
Kinetic energy flowing higher in the atmosphere or the water column is redirected by turbulent mixing onto the turbines in the array \citep*{Calaf2010,Divett2013,Meyers2013}. 
Therefore, subsequent rows of turbines must be placed far enough back for turbulent mixing to redirect kinetic energy from higher elevations and replenish the deficit in the turbine wakes \citep{Newman1977,Lissaman1979,Frandsen1992}. 
The rate of kinetic energy redirection imposes a limit on the performance of turbine arrays, and approaches to overcome this limit are actively sought \citep*{Jimenez2010,Dabiri2011a,Wagenaar2012,Fleming2014,VerHulst2015,Churchfield2015}.

If there are practical limitations on the locations of turbine installations, we can reframe the challenge and focus on redirecting the flow, and therefore more kinetic energy, to those locations.
The flow manipulation we consider changes the oncoming steady freestream to another {\em steady} flow pattern with more power incident on the turbines.
It is not obvious what fundamental fluid dynamic influence underlies such a flow manipulation, what magnitude of redirection would be necessary, or how the flow redirection influences power extraction.
The only mechanism of this type presented in the literature is by yawing axial flow (also known as horizontal axis) turbines to the flow \cite*[see][]{Jimenez2010,Wagenaar2012,Fleming2014,Churchfield2015}.
However, it is unclear how effective the approach of yawing turbines is in relation to any ideal limits on flow deflection that might exist.
In this paper we establish the general feasibility of this type of systematic flow manipulation in two dimensions, provide a simple mathematical model explaining the key features of the flow, derive the characterizing parameters, analyze the idealized limits, and develop general guidelines for designing such an array.

To illustrate the basic fluid mechanical phenomenology accompanying flow deflection, in \S\ref{sec:defvawt} we present two computational examples of ``deflector-turbine'' arrays constrained to be in a narrow strip.
In these cases the turbines are represented by actuator disks of different shapes, and the flow is deflected using either airfoils or a distributed body force on the fluid.
The computational array is subjected to a freestream flow and the power extracted by the array for different deflector strengths is monitored.
In both these cases, the array extracts more power with the wake deflection and imprints an identifiable signature in the surrounding flow structure.
These examples serve to orient the reader toward different ways the flow can be manipulated and guide the various approximations we make in developing approximate theories and reduced-parameter models of flow deflection.
Readers familiar with the basic phenomenology may start reading from \S \ref{sec:defdefturb}.

Next in \S\ref{sec:defdefturb}, we use two-dimensional inviscid fluid dynamics to derive definitions of ``deflectors'' and ``turbines.'' 
A deflector is any fluid dynamic element that introduces bound vorticity but does not shed any free vorticity.
A turbine is any fluid dynamic element that sheds free vorticity (positive and negative amounts at equal rates as required to establish a steady state) into the flow.
Sometimes the same body or machine interacts with the flow to induce bound vorticity and shed free vorticity.
In doing so, it simultaneously act as a deflector and a turbine. 
A deflector-turbine array is a combination of these fluid dynamical element.
A mathematical rationale for these definitions is the first main result of this article.

Next in \S\ref{sec:idealization}, we present a model of a deflector-turbine array in the form of a line segment serving as an internal fluid boundary to form a ``linear deflector-turbine array.''
We impose appropriate boundary conditions on this internal boundary such that the segment interacts with the flow to cause wake deflection and energy extraction in a manner that approximates both the computational cases in \S \ref{sec:defvawt}. 
We investigate the essential features of the resulting flow.
This model and its energy conversion performance presented in \S\ref{sec:results} serve as a common reference for all deflector-turbine arrays that occupy a narrow strip.
The general characterization of energy extraction, including the limits on its performance, by the linear deflector-turbine array and its comparison with the 2-D computational examples is the second main result of this article.
An example application of our analysis to hydrokinetic power is presented in \S\ref{sec:hydrokinetic} and limitations of our analysis are discussed in \S\ref{sec:discussion}.

\begin{figure}
\centerline{\includegraphics[width=0.75\textwidth]{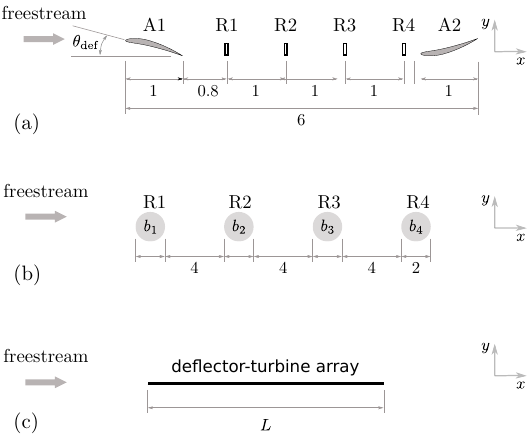}}
\caption{Schematic showing two examples of deflector-turbine arrays and an idealized array. (a) Four rectangular actuator disks of dimensions $0.2 \times 0.05$ labeled R1, R2, R3 and R4 are located unit distance apart along a straight line oriented parallel to the freestream. Two NACA-6409 airfoils, labeled A1 and A2 located upstream and downstream of the these disks. A1 is oriented at an angle of attack $\defaoa$ to the freestream, whereas A2 is oriented upside-down and at an angle of attack $-\defaoa$. All distances are scaled by the airfoil chord. (b) Four model circular turbines of equal size are labeled R1, R2, R3 and R4. Their centers are located 3 diameters apart along a straight line aligned with the freestream. All distances are scaled by the turbine radii. (c) Schematic of an idealized linear deflector-turbine array. The array occupies a segment of length $L$ aligned with the freestream.}
\label{fig:ExampleArraySchematic}
\end{figure}

\section{Example turbine arrays with deflection}
\label{sec:defvawt}
We begin in \S\ref{subsec:deflectors}-\ref{subsec:vawt} with results from two computational examples of deflector-turbine arrays to demonstrate the feasibility and illustrate the basic phenomenology.
These arrays are constructed by combining distributed body forces and/or airfoils, which extract energy and deflect the flow. 
In these examples, schematically shown in Figure \ref{fig:ExampleArraySchematic}~(a-b), we position turbines in a straight line aligned with the freestream flow to highlight the role of flow deflection.
Certain features of these computational models are chosen arbitrarily, e.g., the shapes of the airfoils and the turbines, the parametrization of the force exerted by the turbines, and the mechanism to deflect the flow.
The results obtained depend on these details to an extent.
Despite differences in these example arrays, both share the fluid dynamic signature of simultaneous wake deflection and power extraction, as we show in \S \ref{sec:defdefturb}-\ref{sec:idealization}.
These examples thus motivate and illustrate the general definitions developed in \S \ref{sec:defdefturb} and validate the idealized linear deflector-turbine array presented in \S \ref{sec:idealization}.

The optimal performance of an inline array of actuator disks when the deflection is absent was first derived by \cite{Newman1986} and serves as a useful reference in analyzing the model performances. 
Newman's result states that the maximum value of the combined array power coefficient based on the frontal area of $n$ inline actuator disks is $8n(n+1)/(3(2n+1)^2)$.
For a single turbine ($n=1$), the maximum power coefficient simplifies to $16/27$ in agreement with Betz-Joukowsky (\LBJ) result \citep[see][]{Betz1920,Joukowsky1920,Okulov2012}, and as $n \to \infty$, the coefficient asymptotes to $2/3$.
Newman claimed his calculation to apply only when the turbines are spaced more than half a diameter apart but not so far that turbulent mixing replenishes the deficit in the wake, nor can the row of turbines be too long for the same reason.
In the course of this investigation, we too find that the influence of turbulence could be sub-dominant when turbines are placed close to each other.
In that case, we compare our results with Newman's and find them to be applicable when the deflection is turned off, despite the arbitrarily chosen shapes of our actuator regions.

However, our main motivation is to investigate the performance of the array by deflecting the wake.
In the first example the flow is deflected using airfoils and in the second example it is deflected by a distributed body force motivated by a cross-flow turbine.
With the deflection turned on, the power extracted by the arrays increases due to the additional kinetic energy flux incident on them.
These examples serve to demonstrate the feasibility of increased power generation with flow deflection, provide the motivation and set the stage for the idealized analysis.
Note that we have neglected any externally imposed pressure gradient in the flow, as was done by \cite{Newman1986}.

\subsection{Case of flow deflection by airfoils}
\label{subsec:deflectors}
As a first example, consider a row of turbines between two identical deflectors in the shape of NACA 6409 airfoils subject to a uniform freestream flow, as shown in Figure~\ref{fig:ExampleArraySchematic}~(a).
Quantities are non-dimensionalized such that the airfoils have unit chord, the fluid has unit density, and the freestream flow speed is unity.
The airfoil shapes have a maximum thickness of 9\% of the chord, and a maximum camber equal to 6\% of the chord.
The deflector centers are separated by 5 chords and oriented at an angle of attack of $\defaoa$ anti-symmetrically (see Figure~\ref{fig:ExampleArraySchematic}a).
The rationale for the anti-symmetric arrangement of deflectors will be presented in \S\ref{subsec:mechanistic}.
The turbines are modeled as 0.05 wide $\times$ 0.2 tall rectangular regions R1, R2, R3, and R4 (collectively labeled $R$) spaced uniformly between the deflectors.
The molecular viscosity of the fluid was chosen such that the Reynolds number based on the freestream flow speed and the airfoil chord is $10^5$.
A body force $\bF = -\perm \bu$, where $\perm$ is a parameter, resists the flow in the rectangular regions $R$ simulating the action of turbines extracting energy.
The rate of work done by the fluid against this body force, $\Wdot = \int_R \perm |\bu|^2~\diff A$ is the energy extracted by the turbines.
Tuning the parameter $\perm$ modifies the axial induction factor of the turbines; $\perm=0$ implies that the turbines are turned off.

This system is solved using the $k-\epsilon$ turbulence model \citep*[see][]{Launder1974}
built into the commercial software COMSOL \citep[see the][]{COMSOLCFD2016} for $0\le \perm \le 16$ and $\defaoa$ in the range -3--14$^\circ$.
Both transient solutions starting from potential flow around the airfoils (which develops instantaneously in the start-up) and steady state solutions were computed in a box that approximated an infinite domain\footnote{COMSOL allows setting a maximum time step, which we successively reduce and monitor the change in the extracted power. 
Once the change in the extracted power starts decreasing linearly with the maximum time-step, we conclude numerical convergence and use the solution with the smallest time step in our results.
We also use an analogous method for ensuring convergence with the spatial resolution by successively halving the grid size.}.
A sample transient solution is shown in the Supplementary Material - Movie 1.
In the region of extent comparable to the array dimension, the (mean turbulent) flow approaches a steady state.
The vorticity shed in the wake is expected to meander, but the underlying flow instability is convective in nature. 
For most of the parameters chosen, the unsteady dynamics resulting from the meander occur far downstream of the array and negligibly influences the steady flow near the array.
(For some cases with $\theta_\text{def} = 14^\circ$, the flow did not reach steady state due to boundary layer separating from the airfoils -- a vortex street is shed by the airfoils. 
We characterized the performance of those cases using long-time averaged quantities obtained from a time-resolved unsteady solution of the governing equations.
An averaging interval of $t=160$ was used, which yielded the averages to $<5\%$ accuracy.)
We found the array performance to be insensitive to the use of other turbulence models so long as the boundary layer on the airfoils remains attached (see Appendix \ref{sec:turbmodel} for a discussion on the influence of turbulence).

\begin{figure}
\includegraphics[width=\textwidth]{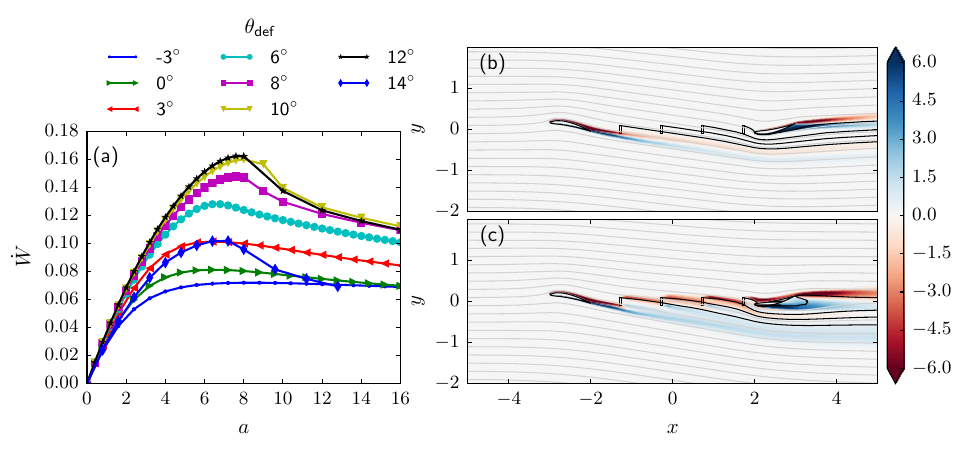}
\caption{(Colour online) Energy extraction rate and sample flow structures resulting in a model of deflector-turbine array using airfoils located at $x=-2.5$ and $x=2.5$, and rectangular actuator disks located at $x=-1.2, -0.2, 0.8, 1.8$. (a) Energy extraction rate $\Wdot$ as a function of $a$ for different angle of attack $\defaoa$ of the deflector airfoils. The data is obtained from steady state solutions, except for the case of $\defaoa=14^\circ$ for which a steady state was not reached because the boundary layer separated from the airfoil. For that case, a time average of a statistically steady transient solution is plotted. (b) Flow structures in the form of streamlines (light gray lines) and vorticity (colour heatmap) for the case $\defaoa=12^\circ$ and $\perm=0$ corresponding to turbines being turned off. Solid black lines show the downstream part of streamlines emanating from the top of the actuator disks. (c) Same as (b) but for $\perm=7.2$.}
\label{fig:Deflector_Turbine_streamlines}
\end{figure}

The power extracted by the turbines from the flow is plotted as a function of $a$ for different deflector angle $\defaoa$ in Figure \ref{fig:Deflector_Turbine_streamlines}~(a).
The case $\defaoa = -3^\circ$ corresponds to the neutral lift orientation for the cambered airfoil shapes we used; we suppose that this case corresponds to no deflection and treat it as the baseline. 
For this case (blue dots), as the parameter $a$ increases, the turbine array extracts more and more power from the flow, exhibiting marginal returns past $a\approx 8$.  
The maximum power generated is about $0.066$, which corresponds to an aggregate efficiency $\eta_\text{agg} = \Wdot/(\frac{1}{2} \rho U^3 A_\text{front})$ of about 0.66, where  $A_\text{front}=0.2$ is the frontal area presented by the turbines to the flow.
This efficiency agrees well with $8n(n+1)/(3(2n+1)^2) = 160/243 \approx 0.658\dots$ for $n=4$, the optimum value of efficiency of multiple inline actuator discs according to \cite{Newman1986}.

To investigate the influence of flow deflection, we increase $\defaoa$ to values greater than $-3^\circ$.
In those cases, we find that an optimum value for $\perm$ appears at which the array extracts more power than the maximum power extracted in the baseline case. 
This maximum power increases as the deflection is made stronger by increasing $\defaoa$ and reaches a maximum $\Wdot=0.16$ at $\defaoa=12^\circ$. 
Nominally, this maximum power corresponds to an aggregate efficiency of $\eta_\text{agg} = 1.6$, however, we show in \S\ref{subsec:comparison} that a different non-dimensionalization provides more meaningful comparison between different deflector-turbine arrays.

An examination of the flow structures sheds light on this behaviour of the array.
Representative ($\defaoa=12^\circ$) flow structures in the form of streamlines and vorticity heatmap that develop are shown in Figure~\ref{fig:Deflector_Turbine_streamlines}~(b-c) for two cases: when the turbines are turned off corresponding to $\perm=0$ and when the array generates maximum power corresponding to $\perm=7.2$.
The airfoils deflect the flow so that it is no longer locally inline with the row of turbines. 
When no energy is extracted (see Figure \ref{fig:Deflector_Turbine_streamlines}~(b)), the streamline grazing the top of the first turbine bypasses most of the area of the second turbine.
For non-zero $\perm$, a wake with a velocity deficit develops behind the actuator disks, most of which is deflected away from the downstream turbines.
With increasing $\defaoa$, the deflection is more and more effective as more undepleted flow is directed onto downstream turbines, and the maximum power extracted increases. 
Above $\defaoa=12^\circ$, the boundary layer on the downstream deflector separates and the airfoils stall.
The deflection mechanism becomes ineffective for larger $\defaoa$ and the array performance deteriorates precipitously.  

\subsection{Case of flow deflection by distributed body force}
\label{subsec:vawt}
Our second model uses a distributed body force localized to a circular region occupied by a turbine (see Figure \ref{fig:ExampleArraySchematic}b).
Quantities are non-dimensionalized such that the circular region has unit radius, the freestream flows with unit speed, and the fluid has unit density.
We apply a body force representing the action of the turbines as
\begin{align}
 f_x = -\perm u + \rot v, \qquad f_y = -\perm v - \rot u,
 \label{eqn:vawtbf}
\end{align}
where $\bF = (f_x, f_y)$ are the Cartesian components of the body force and $(u,v)$ are the components of the fluid velocity field.
The term proportional to $\perm$ acts opposite to the fluid velocity and the term proportional to $\rot$ acts perpendicular to it.
As a result, the power extracted by the array, $\Wdot=\int_R \bF\cdot\bu~\diff A= \int_R \perm |\bu|^2~\diff A$, where $R$ is the region the body force is applied, depends only on $\perm$ and not on $\rot$.
There are several interpretations of the term proportional to $\rot$.
This term does no work on the fluid, but merely changes the direction of the flow, and this is one manner to understand the deflection caused by the body force.
Another interpretation of this term lies in its form being identical to the fictitious Coriolis force that appears in the Navier-Stokes equations when written in a rotating frame, such as is commonly used for geophysical fluid dynamics \citep{Pedlosky2013}. 
The rotation rate of the frame, $\rot$, appears as background vorticity in the region where the body force in \eqref{eqn:vawtbf} acts.
This vorticity acts as bound vorticity in the turbines, imparts curvature to the streamlines, and deflects the flow.
The rotation rate, and the accompanying deflection of the flow, is turned off by choosing $\rot=0$, while the turbines may be turned off by setting $\perm=0$.

The array is composed of four turbines with centers located at $x$ = 0, -6, -12 and -18, which corresponds to $A_\text{front}=2$.
The parameter $a$ for all four turbines is equal, but the rotation rates are not.
The rotation rates of the four turbines $\rot_1$, \dots $\rot_4$ are chosen such that $\rot_1 = 4\rot_2 = -4 \rot_3 = -\rot_4 = 4 \beta$, where the parameter $\beta$ is varied to influence the strength of flow deflection.
The rationale behind this choice for the rotation rates, especially the anti-symmetric arrangement, is presented in \S\ref{subsec:mechanistic}.
We use COMSOL's Computational Fluid Dynamics module and solve the Navier Stokes equations with the $k-\epsilon$ model for turbulence starting from a uniform flow.
The molecular viscosity is chosen such that the Reynolds number based on the turbine radius and the freestream is $10^6$.
The transient evolution of the flow is shown in the Supplementary Material - Movie 2.
The flow approaches a steady state in the region around the array, which we further examine here.

\begin{figure}
\includegraphics[width=\textwidth]{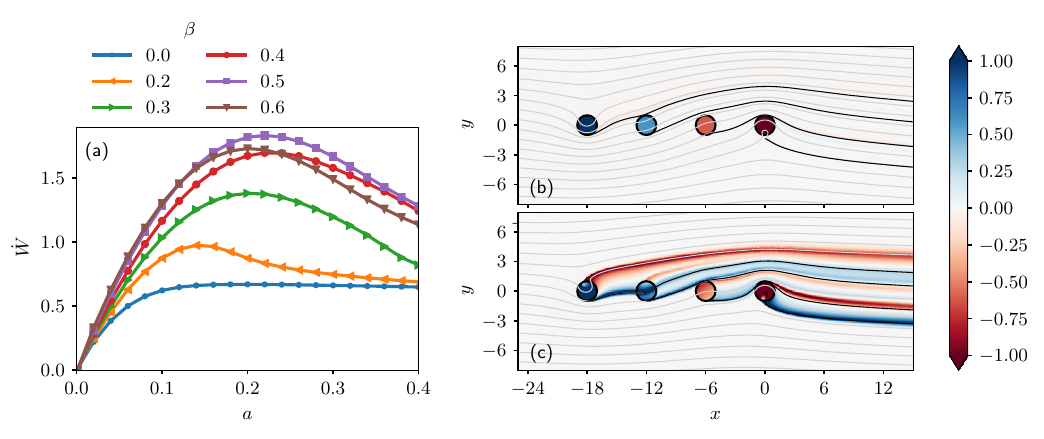}
\caption{(Colour online) Energy extraction rate and sample flow structures resulting in a model deflector-turbine array described in \S\ref{subsec:vawt}. (a) Energy extraction rate $\Wdot$ as a function of $a$ for different $\beta$. (b) Flow structures in the form of streamlines (light gray lines) and vorticity (colour heatmap) for the case $\beta=0.6$ and $\perm=0$ corresponding to turbines being turned off. Solid black lines show the downstream part of streamlines passing through the bottom of the turbines. (c) Same as (b) but for $\perm=0.19$.
} 
\label{fig:VAWT-streamlines}
\end{figure}

The power extracted by the array in steady state is shown in Figure~\ref{fig:VAWT-streamlines}~(a).
The case $\beta=0$, where the deflection is turned off, acts as the baseline case. 
In this case, the power extracted by the array monotonically increases with $\perm$ and then asymptotically approaches a maximum $\Wdot = 0.66$ for large $\perm$. 
This value corresponds to an $\eta_\text{agg} = 0.66$, which is consistent with the maximum power extracted by a row of four actuator disks according to \cite{Newman1986}.
As the deflection is strengthened by increasing $\beta$, the maximum extracted power occurs for a finite $\perm$ in the range 0.17-0.24
\footnote{Note that the differences between \S\ref{subsec:deflectors} and \S\ref{subsec:vawt} in values of $\perm$ where the maximum power occurs is due to the differences in geometry.
The body force in this example acts over a larger region, and therefore comparable effect is achieved at a lower value of the coefficient.}, and the maximum value of $\Wdot$ increases with $\beta$.
This increase saturates for $\beta\gtrsim 0.5$, the maximum power extracted is $\Wdot=1.83$, corresponding to the non-dimensional $\eta_\text{agg} = 1.83$.
Again, $\eta_\text{agg}$ is not the best non-dimensional representation of the array performance, and we show in \S \ref{subsec:comparison} that a different non-dimensionalization allows for a better comparison between different deflector-turbine arrays.

An examination of the flow structure helps explain these observations.
The streamlines and a vorticity heatmap for $\beta=0.6$, and $\perm=0$ and $\perm=0.19$ are shown in Figure \ref{fig:VAWT-streamlines}~(b-c) respectively. 
When the turbines are turned off, the vorticity is confined to the regions of the turbines and no vorticity is shed in the flow in steady state.
In fact, the vorticity in each turbine is equal to its value of $\rot$.
Associated with this vorticity is the curvature of the streamlines and the deflection of flow as seen in Figure \ref{fig:VAWT-streamlines}~(b).
The deflection for $\beta=0.6$ is strong enough such that the streamline passing through the bottom extreme of the upstream turbines grazes past the top extreme of the next downstream turbine.
For non-zero $\perm$ this flow diverts the wake of upstream turbines to bypass the downstream ones, thereby redirecting undepleted flow to the turbines and improving the performance of the array.
There is no analog in this case of boundary layer separation that occurs on airfoils in \S\ref{subsec:deflectors}, however another mechanism limits the maximum power in this case.
The wake is almost completely deflected by the flow for the case of $\beta\approx 0.6$ and any stronger deflection does not increase the flux of kinetic energy incident on the turbines.
Consequently, the maximum extracted power also stops increasing with increase in $\beta$.

We note that this model for turbines is inspired by cross-flow turbines, also known as vertical axis turbines in the context of wind power.
The parameter $\rot$ represents the bound vorticity arising due to the rotation of such turbines.
In cross-flow turbines the induction factor and the rotation rate are coupled, implying that $\perm$ and $\rot$ are not independent of each other. 
Here we do not explore the relationship between our model and cross-flow turbines any further. 
Nevertheless, this example array represents another possible flow manipulation, in this case using a force field, that extracts power and deflects the wake.
Whether a body force may represent cross-flow turbines, while immaterial for the discussion in this article, is interesting from the perspective of reduced-parameter models of cross-flow turbines.

The simulated airfoil and distributed-body-force examples demonstrate that deflection can improve the power extraction within small arrays. 
We show in Appendix \ref{sec:turbmodel} that the influence of turbulence in our numerical solutions is negligible.
If the influence of turbulence could similarly be neglected in practice, then the essential features of the flow would be well-captured by inviscid fluid dynamics. 
In order to further understand the interaction between wake deflection and power extraction and limits of such an array's utility, we turn to idealized modeling. 

\section{Definition of deflectors and turbines}
\label{sec:defdefturb}
Next we present an idealization to approximate and unify all flow manipulations that extract power from the flow and deflect the wake away from the array, as exemplified by the two cases in the previous section.
Our idealization neglects the viscosity and turbulent processes, and simplifies the framework to two-dimensional inviscid fluid dynamics.
The deflector-turbine array is assumed to occupy a bounded, but otherwise arbitrary, area in the two-dimensional plane, where it may exert forces on the fluid.
Using this framework, we seek to derive appropriate definitions for ``deflectors'' and ``turbines''.
A detailed discussion on the role of wake recovery by turbulent mixing, which is the predominant influence of viscosity for flows with a large Reynolds number, is presented in \S\ref{sec:hydrokinetic} for an application to hydrokinetic power.

Our analysis proceeds from the Biot-Savart law relating fluid velocity to vorticity; the incompressible flow velocity $\bu(\bx)$ at any point $\bx$ in the two-dimensional $x-y$ plane (with $z$ perpendicular to the plane) around the array may be described in terms of the vorticity $\vort(\bx) \zhat = \grad \times \bu(\bx)$ as
\begin{align}
 \bu(\bx) = U\xhat + \int_A \dfrac{\omega(\bx') \zhat \times (\bx-\bx')}{2\pi |\bx-\bx'|^2}~\diff A',
\label{eqn:biotsavart}
\end{align}
where $\xhat$ and $\zhat$ are unit vectors in the $x$ and $z$ direction respectively, and the integration is carried out over $A$, the whole two-dimensional $x-y$ plane. 
The vorticity may be decomposed in terms of bound vorticity $\vortb$ and the free vorticity $\vortf$ as $\vort = \vortb + \vortf$, defined as follows.
Bound vorticity is vorticity field present in a region where the array in principle could exert a force on the fluid.
Free vorticity is the vorticity field that obeys Helmholtz law in the region where the array in principle cannot exert any force on the fluid.
In the framework of inviscid dynamics the most general influence the array may have on the flow can be represented through the bound and free vorticity.
In this way, vorticity dynamics may be used to infer the most general manipulation the deflector-turbine array can have on the flow.
Therefore, ``deflectors'' and ``turbines,'' as we defined them, represent the most general manipulations the array can have on the flow.

\subsection{Flow deflection and bound vorticity} 
\label{subsec:bdvorticity}
Here we show that in a compact, but otherwise arbitrary, region occupied by the array bound vorticity alone cannot extract power.
Therefore, in our framework, the bound vorticity is a fictitious instrument (in the most general case agnostic to the method of fluid deflection) that represents the effect of the deflector array. 
This result follows from the observation that a velocity field influenced only by bound vorticity recovers to its freestream value in the far-field.

In the absence of any free vorticity, and by our definition of bound vorticity, $\vort(\bx')$ in \eqref{eqn:biotsavart} is non-zero only in a compact region occupied by the array.
The far-field velocity due to the bound vorticity distribution can then be written by substituting $|\bx| \gg |\bx'|$ in \eqref{eqn:biotsavart} and simplifying to get 
\begin{align}
\begin{split}
 \bu(\bx) \sim U\xhat &+ \dfrac{\Gamma_1}{2 \pi |\bx|} \thetahat + O\left( \dfrac{1}{|\bx|^2} \right)
\end{split}
\label{eqn:ubound}
\end{align}
where $\Gamma_1 = \int_{A} \omega(\bx') dA'$ is the net bound vorticity, and $\thetahat$ is the unit vector along the azimuthal direction in a polar coordinate system centered at the centroid of the bound vorticity distribution. The corresponding flux of kinetic energy is
\begin{align}
 \dfrac{1}{2} \rho |\bu|^2 \bu \sim \dfrac{1}{2} \rho \left[U^3 \xhat + \dfrac{U^2\Gamma_1 }{2 \pi |\bx|} \left( \thetahat + 2 \xhat (\xhat\cdot\thetahat) \right) + O\left( \dfrac{1}{|\bx|^2} \right)\right].
 \label{eqn:keflux}
\end{align}
Similarly, the pressure $p$ may be found using Bernoulli equation for potential flow $p + \rho |\bu|^2/2 = p_\infty + \rho U^2/2$, where $p_\infty$ is the pressure far from the array.
Substituting \eqref{eqn:ubound} in Bernoulli equation yields
\begin{align}
 p - p_\infty = -\dfrac{\rho U \Gamma_1}{2 \pi |\bx|} \xhat \cdot\thetahat + O\left( \dfrac{1}{|\bx|^2} \right).
\end{align}
The quantity $(p-p_\infty)\bu$ represents the work done by pressure, which has the asymptotic expansion
\begin{align}
 (p-p_\infty) \bu = -\dfrac{\rho U^2 \Gamma_1}{2 \pi |\bx|} (\xhat \cdot\thetahat) \xhat + O\left( \dfrac{1}{|\bx|^2} \right),
\label{eqn:peflux}
\end{align}
for large $|\bx|$.

The energy extracted by the array can be found by a contour integral of the normal component of the kinetic energy flux and the work done by pressure along any closed curve $C$ bounding the array, which may be written as
\begin{align}
\begin{split}
\Wdot &= \oint_C \left[ \dfrac{1}{2} \rho |\bu|^2 + (p-p_\infty) \right] \bu\cdot \bn~\diff s \\
      &= \oint_C \dfrac{\rho U^2}{2} \left[ U \xhat + \dfrac{\Gamma_1 }{2\pi |\bx|} \thetahat  + O\left( \dfrac{1}{|\bx|^2} \right) \right] \cdot \bn~\diff s.
\end{split}
\label{eqn:defpowerbalance}
\end{align}
where $s$ represents the arc-length coordinate along $C$, and $\bn$ the unit normal to $C$.
In arriving at the second line of \eqref{eqn:defpowerbalance}, the curve $C$ is assumed to be much larger than the array size, so that approximations \eqref{eqn:keflux} and \eqref{eqn:peflux} may be used.
If the contour is chosen to be a circle of radius $R_c$ centered at the origin, then the normal to the curve is along the unit radial vector $\rhat$, and the resulting terms $\xhat \cdot \rhat$ and $\thetahat\cdot\rhat$ integrate exactly to zero.
Therefore, the leading order contribution to the energy deficit integral is at most $O(1/R_c)$.
Since the value of this integral is independent of the specific contour $C$, as long as it encloses the array, taking the limit $R_c\to\infty$ shows that the remaining terms also integrate to zero.
The wake of the array therefore has no energy deficit, and the array cannot extract any energy.

The bound vorticity, however, alters the fluid flow, changing the rate of fluid volume, momentum and kinetic energy incident on the array
The bound vorticity can be considered to be the fluid dynamical quantity that determines flow deflection without energy extraction.
Therefore, the distribution of bound vorticity can be treated as the primary design variable that influences flow deflection in arrays.
We define ``deflectors'' as elements that introduce bound vorticity into the flow.



\subsection{Energy extraction and free vorticity}
In the framework presented in \S\ref{sec:defdefturb}, vorticity is shed at equal and opposite rates from each turbine.
(Note that for a steady state flow of an inviscid fluid the net rate of free vorticity shedding must be zero.)
Part of the shed vorticity either passes through, grazes past or bypasses the downstream turbine.
In the absence of an external force on an inviscid fluid, the free vorticity in two-dimensions is simply advected by the fluid without any change in its strength.
Therefore, the strength of this vorticity changes only if the streamline carrying it passes through or grazes past other turbines or deflectors.
Once the streamline exits the array, the shed vorticity strength remains unaltered along the streamline \citep{Saffman1992,Lighthill1986}, and in this manner influences the wake far downstream of the array, where the flow becomes parallel to the freestream.
The flow velocity there may be approximated as $u=U_w(y)$, and $v=0$ for some profile $U_w(y)$.
The free vorticity profile far downstream is related to $U_w(y)$ as 
\begin{align}
\vortf = - \dfrac{\diff U_w}{\diff y}.
\label{eqn:vortwakedef}
\end{align}
Note that $U_w = U$ away from the wake, which is the region outside the influence of the array, where $\vortf=0$.
Equation \eqref{eqn:vortwakedef} shows that the wake may have a deficit (or an excess), defined as the difference between $U_w(y)$ and $U$, if and only if the free vorticity is non-zero.
If power were extracted from the flow, then this wake must have a commensurate kinetic energy deficit, and therefore a velocity deficit.
Thus, the kinetic energy deficit is a direct consequence of the free vorticity shed from the turbines in the array.
Therefore, ``turbines'' are defined as elements that introduce free vorticity in the flow.

Note that in the example presented in \S\ref{subsec:deflectors}, free vorticity is also shed from the boundary layers attached to the plate, which contributes to the wake deficit and power removed from the flow by viscous dissipation.
Thus, in this framework, there is no difference between useful power extracted by the turbines and power loss resulting from parasitic effects.
A further direct consequence of the velocity deficit in the wake and conservation of fluid mass is a lateral expansion of the stream tube.
This stream tube expansion, is also visible in Figures \ref{fig:Deflector_Turbine_streamlines}~(c) and \ref{fig:VAWT-streamlines}~(c) for the two examples respectively.


\section{Idealization as a linear deflector-turbine array}
\label{sec:idealization}
We now seek a simplified description of the performance of the two example arrays presented in \S\ref{subsec:deflectors}-\ref{subsec:vawt}. 
We begin by noting the common features between the example arrays based on results of \S\ref{sec:defdefturb}, which exist despite the apparent differences in the shape of turbines and the deflection mechanism.
In both cases, the deflection mechanisms induce a non-uniform and non-trivial distributions of bound vorticity,
In the first example in \S\ref{subsec:deflectors}, the bound vorticity is established by the presence of the deflector plates and their angle of attack to the flow.
In the second example in \S\ref{subsec:vawt}, the bound vorticity is set up by the Coriolis-like body force acting within the model turbines, which also deflects the flow.
The common feature of the bound vorticity in both cases is that the resulting streamlines passing through upstream turbines are deflected to partially or completely bypass the downstream ones.
Similarly, in both cases, the body force resisting the flow leads to a complicated profile of free vorticity shed in the flow.
Most of the vorticity is shed near the extreme ends of individual turbines, and part of it is modified as it passes through downstream turbines.
However, in both cases the signature of collective energy extraction is encoded in the profile of free vorticity outside the array, which ultimately causes the wake deficit.
Finally, we note that both the example arrays have a long slender geometry. 
The array length in the streamwise direction for the first array is larger than its width (cross-stream dimension) by a factor 20.
For the second array this factor is 10.
Due to the large aspect ratio, and in case where the flow deflection is effective, the kinetic energy flux incident on the frontal area may be neglected in comparison to the deflected flux.
Guided by these observations, we idealize the geometry of the array and the shapes of the bound and free vorticity profiles to arrive at the following ``linear deflector-turbine array''.

Consider the uniform two-dimensional flow of an inviscid fluid of density $\rho$ in an infinite domain with freestream velocity $U$ oriented parallel to the $x$ axis (see Figure~\ref{fig:ExampleArraySchematic}c).
The deflector-turbine array is represented by a segment of length $L$ centered at the origin and oriented along the freestream. 
The performance of the array is measured in terms of the power $\Wdot$ extracted by the array, rendered dimensionless using the three parameters ($\rho$, $U$ and $L$) in the statement of the problem as
\begin{align}
 \Ca = \dfrac{\Wdot}{ \frac{1}{2}\rho U^3 L}.
\label{eqn:nondim}
\end{align}
We call $\Ca$ the array power density.

If the deflectors in the idealized array were turned off, the analysis by \cite{Newman1986} would apply and predict that at most 2/3 of the kinetic energy flux incident on the frontal area could be extracted.
However, due to our geometric idealization, our linear deflector-turbine array poses zero frontal area to the flow, and therefore zero kinetic energy flux incident on itself.
Such an array, according to \cite{Newman1986}, could not extract any power from the flow.
However, mathematically the line segment appears as an internal boundary of the domain and may influence the flow in a more general way than represented by Newman's analysis.
Based on our analysis in \S\ref{sec:defdefturb}, the most general influence the linear array may have on the flow may be represented through the bound vorticity within the array and the free vorticity shed form the array.

%
We do not seek to derive the exact distributions of the bound and shed vorticity profile for the two cases in \S\ref{subsec:deflectors}-\ref{subsec:vawt}
Instead, in the spirit of retaining the least parameters, here we seek to identify simple profiles for bound and shed vorticity that capture the essential features of the array's influence on the flow.
Therefore, we interpret the linear deflector-turbine array as a reduced parameter model for any deflector-turbine array occupying a long narrow strip.
The usefulness of our idealization will be evaluated based on its ability to explain the performance of both the examples in \S\ref{sec:defvawt}.

\subsection{Choice of bound vorticity profile and the resulting deflected flow}
\label{subsec:ldta-bd}
We choose the bound vorticity in the form of a vortex sheet, and choose its strength such that the component of velocity normal to the array is uniform along the length of the array.
The magnitude of the normal component $v$ is chosen such that $v = U \tan \aoa$, for a parameter $\aoa$ describing the deflection.
The strength of the bound vortex sheet can be determined using complex analysis (see Appendix \ref{sec:complexsol} for details) to be 
\begin{align}
\vortb(\bx) = \vortbo(x) \delta(y) = -\dfrac{2U x \tan \aoa}{\sqrt{L^2/4 - x^2}} \delta(y), 
\label{eqn:boundvorticity}
\end{align}
where the array is located at $-L/2<x<L/2$ and $\delta(y)$ is the Dirac-delta function.
An example of the resulting flow field is shown in Figure \ref{fig:Summary1}~(a) for $\aoa = 40^\circ$.
Note that the component of velocity tangential to the sheet is discontinuous across the array by an amount $\vortbo(x)$.

The bound vorticity deflects fluid at the rate $Q_0$ to flow through the array.
Two recirculation regions develop near the ends of the array.
The size of the recirculation region overlapping with the array is characterized in detail in the Appendix \ref{sec:recirclen}.
The dependence of $Q_0$ on $\aoa$ can be calculated analytically (see Appendix \ref{sec:complexsol} for details), expressed in dimensionless terms as 
\begin{align}
\dfrac{Q_0}{UL} = \zeroflow,
 \label{eqn:zeroflow}
\end{align}
and is plotted in Figure \ref{fig:FlowPower}~(a).

The parameter $Q_0/(UL)$ represents the dimensionless ``deflected-flow strength'', and parameterizes the effectiveness of flow deflection.
This dependence has two asymptotic limits; for $\tan \aoa\ll 1$, which we term the weak deflection limit, $Q_0/(UL) \approx \tan\aoa$ and for $\tan \aoa \gg 1,$ which we term the strong deflection limit, $Q_0/(UL) \approx \sqrt{\tan\aoa}$.
These limits may be understood by considering the following: The flow rate $Q_0$, is set by the product of the velocity component normal to the array, $U\tan\aoa$, and the length $\ell_0$ of the region between the recirculation zones, $Q_0= U\ell_0 \tan\aoa $.
For weak deflection, the recirculation regions are small, and consequently $\ell_0 \approx L$, therefore leading to $Q_0 \approx UL\tan\aoa$. 
While for strong  deflection, the recirculation region occupies almost all of the array length leading to $\ell_0 \approx L/\sqrt{\tan\aoa}$, implying $Q_0 = UL\sqrt{\tan \aoa}$.
These two asymptotes are also shown in Figure \ref{fig:FlowPower}~(a).

As we showed in \S\ref{subsec:bdvorticity}, with this bound vorticity alone, the wake of the array has no energy deficit, and the array cannot extract any energy.
Therefore, we formally define $Q_0$ to be the flow rate incident upon the array in the absence of energy extraction.
This deflection of the flow implies that kinetic energy is incident on the array at the rate $\rho Q_0 U^2/2$.
We show next that by shedding an appropriate profile of free vorticity, a fraction of this incident energy can be extracted from the flow.

\subsection{Choice of shed vorticity profile and the resulting wake profile}
A representative example showing the profile of shed vorticity we choose and the resulting flow is shown in Figure \ref{fig:Summary1}~(b).
We choose to shed free vorticity such that the far downstream wake formed by the deflector-turbine array has a uniform speed $V\le U$ and the widest possible cross section.
We achieve such a wake by shedding vorticity with strength $\pm\Omega$ at points labeled $s_1$ and $s_2$ shown in Figure \ref{fig:Summary1}~(b) along the extreme streamlines passing through the array without recirculating.
A consequence of this choice and inviscid vorticity dynamics is that the shed vorticity is advected along the streamline and forms a vortex sheet.
We determine the location of $s_{1,2}$ and the shape of this vortex sheet numerically, as described in \S\ref{sec:nummeth}.

Applying energy balance to a control volume, shown as ABCD in Figure \ref{fig:Summary1}~(b), between the two streamlines on which vorticity is shed, yields the deficit of kinetic energy flux in the wake as
\begin{align}
\Wdot = \dfrac{1}{2} \rho Q (U^2 - V^2),
\label{eqn:kedeficit}
\end{align}
where $Q$ is the modified flow through the array to be determined numerically.
This expression may also be derived by applying energy conservation across the array to determine the distribution of power extracted per unit length, as is presented in \ref{sec:powerperunitlength}.
Conservation of energy implies that for inviscid flow the power extracted by the array is limited by this deficit.
The flow rate with energy conversion, $Q$, is less than the value without energy conversion, $Q_0$, because the turbines impart additional resistance to the flow and change the bound vorticity. 
We refer to the ratio $Q/Q_0$ as the ``flow reduction factor''.

Far downstream, where the streamlines become parallel to the freestream, a consequence of the relation \eqref{eqn:vortwakedef} between $\omega$ and $U_w$ is that the shed vortex sheet has strength $\Omega = U-V$.
Through this relation with $V$, the aggregate turbine induction factor of the array can be tuned by varying $\Omega$.
The expansion of the stream tube passing through the array is also visible in Figure \ref{fig:Summary1}~(b).

The dimensionless quantities $Q_0/(UL)$ (or equivalently $\aoa$ through \eqref{eqn:zeroflow}) and $V/U$ parameterize our choice of bound and free vorticity distributions.
In the subsequent analysis, we determine their influence on the array performance.
The deflection of the streamlines to direct fluid through the array, the deficit in the wake, and the expansion of the streamlines are the essential flow features our model represents and follow directly from our choice of bound and shed vorticity.

\subsection{Numerical method for determining the total vorticity field}
\label{sec:nummeth}
We numerically determine the streamlines, the bound vorticity, and the shape of the free vortex sheet in a self-consistent manner such that the free vorticity is constant along the streamlines and the flow velocity maintains an angle of attack $\aoa$ with the freestream flow direction.
The bound vorticity is localized to the turbine array $-L/2<x<L/2$, and described as $\vortb = \vortbo(x) \delta(y)$.
The free vorticity is represented in the form of sheets of strength $\Omega = (U-V)$ along two streamline curves $\bx_n(s)$, $n=1,2$, so $\vortf(\bx) = \Omega (\delta(\bx-\bx_1) - \delta(\bx-\bx_2))$, where $0<s<\infty$ parameterizes the curves.
The Biot-Savart integral for stream function in terms of $\vortbo$, $\bx_1$ and $\bx_2$ may be written as
\begin{align}
\begin{split}
\psi(\bx; \bx_1, \bx_2, \vortbo) = -Uy + \int_{-L/2}^{L/2} \dfrac{\vortbo(s) \log |\bx-s\xhat|}{2\pi}~\diff s \\
         + \Omega \int_0^\infty ~\dfrac{ \log |\bx-\bx_1|~x'_1(s) - \log |\bx - \bx_2 |~x'_2(s)}{2\pi}~\diff s  
\end{split}
\label{eqn:biotsavartpsi}
\end{align}
where substituting the localized forms of $\vort$ converts the area integral \eqref{eqn:biotsavart} to line integrals.
We numerically approximate this turbine array and the bound vorticity integral by discretizing along Chebyshev points $x'_k = L\cos\theta_k/2$, where $\theta_k = \pi k/N$, $k=0,1,\dots, N$, and half-Chebyshev point corresponding to $k = 1/2, 3/2, \dots, N-1/2$.
On the half-Chebyshev points, $\vortbo(x'_{k-1/2}) = \vortbo_{k-1/2}$.
The stream function at the half-Chebyshev points $x'_{j-1/2}$ due to the bound vorticity is represented as a matrix-vector product using a matrix $M_{jk}$ as
\begin{align}
\tilde\psi_{j-1/2} = \int_{-L/2}^{L/2} \dfrac{\vortbo(s) \log |x'_{j-1/2}-s|}{2\pi}~\diff s \approx \sum_{k=1}^{N} M_{jk} \vortbo_{k-1/2}, \qquad j=1,2,\dots,N.
\end{align}
The matrix $M_{jk}$ is spectrally approximated and determined numerically through the set of linear maps
\begin{align}
 \tilde\psi_{j-1/2} &= \sum_{n=1}^N a_n \cos n\theta_{k-1/2}, \\
 \text{where $a_n$'s satisfy} \quad \vortbo_{k-1/2} &= \dfrac{4}{L \sin \theta_{k-1/2} } \sum_{n=1}^N n a_n \cos n\theta_{k-1/2}.
\end{align}

The Biot-Savart line integral \eqref{eqn:biotsavartpsi} for free vorticity is evaluated by discretizing the curve $\bx_n$ in to $N_n$ linear segments.
The total integral is a sum of the contributions from individual segments.
The integral over individual segments can be evaluated analytically, as shown in Appendix \ref{sec:biotsavartsegment}.
%
The stream function integral from the bound vorticity is also evaluated for points outside the turbine array by approximating $\vortbo$ to be a constant $\vortbo_{k-1/2}$ on each segment between the Chebyshev points $x'_{k-1}$ and $x'_k$.

The mean velocity components tangential and normal to the array are
\begin{align}
u = \lim_{\epsilon \to 0} \dfrac{\psi(x, \epsilon) - \psi(x, - \epsilon)}{2\epsilon}, \quad
v = \dfrac{\partial \psi}{\partial x},
\end{align}
where the limit is approximated numerically using a small value $10^{-3}$ for $\epsilon$.
The limiting process described for evaluating $u$ yields the mean value of the velocity component tangential to the array; such a process is necessary because the tangential velocity undergoes a discontinuous transition due to the presence of a bound vortex sheet in the deflector-turbine array.

The discretized bound vorticity $\vortbo_{k-1/2}$, $k=1,2,\dots,N$, is chosen such that a constant angle is maintained between the average velocity vector and the array, in other words, $v = u \tan \aoa$ at the half-Chebyshev points.
This criteria does not uniquely determine the bound vorticity distribution but instead leads to a one-parameter family parameterized by the circulation $K$ \citep{Batchelor2000}; an amount $K/\sqrt{L^2/4-x^2}$ added to the bound vorticity for arbitrary $K$ does not change the flow direction at the deflector array.
We choose the $K$ which leads to zero circulation in the far-field; we have empirically found our results to be insensitive to other choices of $K$.

Free vorticity strength is constant along streamlines, implying
\begin{align}
 \psi(\bx_k; \bx_1, \bx_2, \vortbo) = \text{constant}, \qquad k=1,2.
 \label{eqn:consistency}
\end{align}

The self-consistent free vortex sheet is determined iteratively starting from an initial guess $\bx_k^0$ and subsequent iterates determined as follows.
The free vortex sheet is defined by the most extreme streamlines that pass through the array and do not recirculate.
The stream function is evaluted on a fixed Cartesian grid and GNU Octave's \texttt{contour} function \citep*{Eaton2014} is used to find an intermediate guess for the streamlines $\tilde{\bx}_k^{q+1}$,
\begin{align}
 \psi(\tilde{\bx}_k^{q+1}; \bx_1^{q}, \bx_2^{q}, \vortbo^{q}) = \text{constant}, \qquad k=1,2, \quad q=1,2,\dots.
\end{align}
The constant is chosen using a bisection algorithm by detecting whether the streamlines form closed contours or extend to far downstream to determine streamlines bounded by the recirculation region at the turbine array ends.
A fractional update $\bx_n^{q+1} = \sigma \tilde{\bx}_n^{q+1} +(1-\sigma) \bx_n^q$, with $0<\sigma \le 1$, is used to ensure convergence.

\section{Results: Limits on $\Ca$}
\label{sec:results}

\begin{figure}
\includegraphics[width=0.98\textwidth]{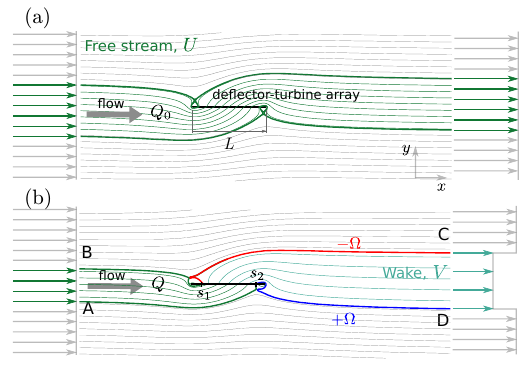}
\caption{(Colour online) Example of flow around a linear deflector-turbine array. 
(a) The ideal flow field arising from a linear deflector-turbine array, such that the flow is deflected to flow through the array at $\aoa=40^\circ$. 
The flow corresponds to an array deflecting the flow with no energy extraction.
$Q_0$ is the deflected-flow rate between the two extreme streamlines (bold green) that pass through the array.
(b) Same as (a) but with energy conversion.
The flow through the array is now reduced from $Q_0$ to $Q$, and the flow speed in the wake is reduced from the freestream $U$ to $V$. 
The boundary of this stream tube constitutes two vortex sheets with strength $\pm \Omega$, shown in blue and red respectively, where $\Omega = U-V$.
The vortex sheets result from free vorticity shed by the array at points $s_1$ and $s_2$.
}
\label{fig:Summary1}
\end{figure}

\begin{figure*}
\centerline{
\includegraphics[width=\textwidth]{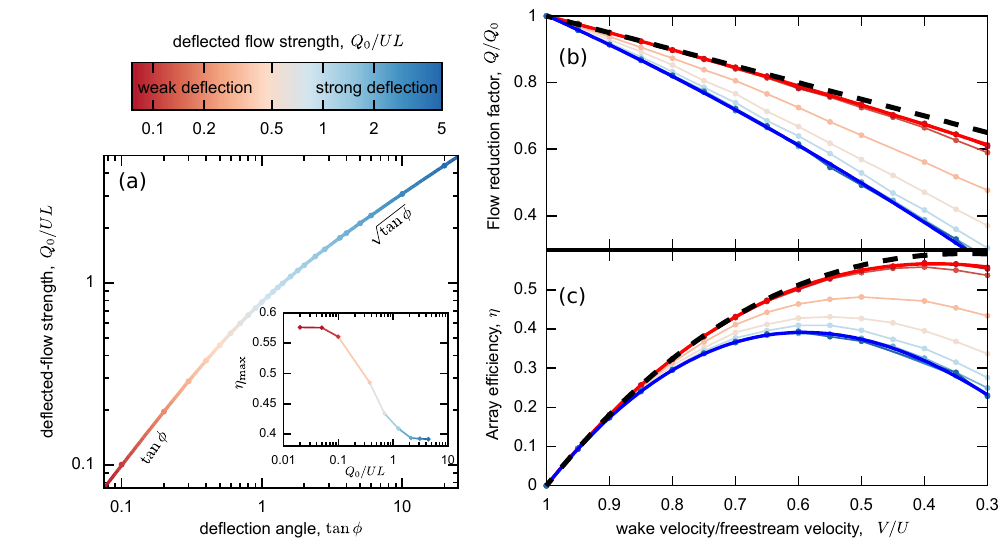}
}
\caption{ (Colour online)
Results on deflector-turbine array performance from the solution of inviscid flow. 
(a) The dependence of the deflected-flow strength, $Q_0/(UL)$, on the deflection angle $\aoa$ calculated in \eqref{eqn:zeroflow}. 
Inset shows the maximum array efficiency possible for a given deflected-flow strength.
The color shown in the palette encodes the deflected-flow strength in this figure.
(b)~The flow reduction factor, $Q/Q_0$, and (c) array efficiency, $\eta$, as a function of the dimensionless wake speed, $V/U$, for different values of the deflected-flow strength encoded by the color.
The dashed black curve shows the performance, according to the \LBJ~analysis, if the entire array was a single actuator disk oriented perpendicular to the incident flow.
The solid red (topmost) and blue (bottommost) curves represent the asymptotic limits corresponding to weak and strong deflection introduced in \eqref{eqn:asympdefs}. 
The curves with symbols correspond to values of $Q_0/(UL)$ of 0.02 (topmost), 0.05, 0.1, 0.2, 0.4, 0.84, 2.0, 5.0, 10.0, and 20.0 (bottommost).
}
\label{fig:FlowPower}
\end{figure*}

Having computed the flow numerically using the aforementioned method, we next present the dependence of the dimensionless array power density on flow parameters.
Using \eqref{eqn:nondim} and \eqref{eqn:kedeficit}, this dependence can be separated into the influence of flow deflection and energy extraction as
\begin{align}
\Ca &= \dfrac{Q_0}{UL} \eta,  \label{eqn:ca} \\
\text{where} \qquad   \eta &\equiv \dfrac{\Wdot}{\rho Q_0 U^2/2} = \dfrac{Q}{Q_0} \left( 1- \dfrac{V^2}{U^2}\right)
\label{eqn:arrayeff}
\end{align}
is the array efficiency, defined as the fraction of the impinging kinetic energy the array extracts. 
Here, we explain why increasing $\aoa$ strongly increases the deflected flow strength, $Q_0/(UL)$,  while slightly reducing the array efficiency, $\eta$ (see Figure \ref{fig:FlowPower}).

\subsection{Array efficiency $\eta$}
We next determine how $\eta$ depends on the flow parameters.
According to \eqref{eqn:arrayeff}, $\eta$ depends on $Q/Q_0$, the flow reduction factor, and $V/U$, the dimensionless wake speed.
The resulting flow reduction factor, $Q/Q_0$, and the array efficiency, $\eta$, are plotted as a function of $V/U$ for fixed $Q_0/(UL)$ in Figure \ref{fig:FlowPower}~(b).
As $V/U$ decreases below unity, and more energy is extracted by the array, $Q/Q_0$ decreases monotonically below unity. 
Each line on Figure \ref{fig:FlowPower}~(b) represents a different deflected-flow strength.
The flow reduction factor, $Q/Q_0$ decreases monotonically with increasing deflection from the limiting value for weak deflection $f_0$ to that of strong deflection $f_\infty$.  The two limits are
\begin{align}
\begin{split}
\text{Weak deflection:}  & \quad \dfrac{Q}{Q_0} \to f_0     \left(\dfrac{V}{U} \right) \text{ for } \dfrac{Q_0}{UL} \ll 1, \\
\text{Strong deflection:}& \quad \dfrac{Q}{Q_0} \to f_\infty\left(\dfrac{V}{U} \right) \text{ for } \dfrac{Q_0}{UL} \gg 1.
\end{split}
\label{eqn:asympdefs}
\end{align}
We use a consistent color scheme in Figure \ref{fig:FlowPower} to represent weak and strong deflection limits in red and blue respectively.

\begin{figure}
\centerline{\includegraphics[width=0.5\textwidth]{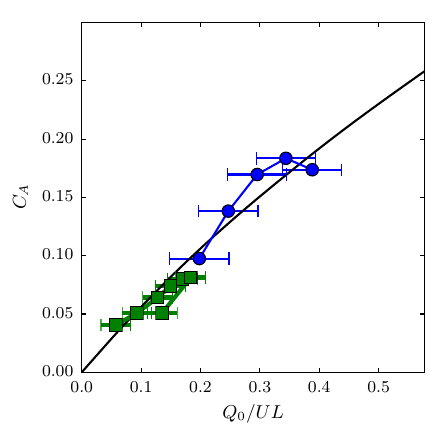}}
\caption{(Colour online) The maximum dimensionless power $\Ca$ from \eqref{eqn:ourlimit} extracted by the idealized deflector-turbine array (solid line) as a function of the deflected-flow strength $Q_0/(UL)$. 
Also plotted are the maximum $\Ca$ observed in the examples using deflector airfoils in \S\ref{subsec:deflectors} (green squares) and a body force in \S\ref{subsec:vawt} (blue circles).
The flow incident on the frontal area of the example arrays constitutes our uncertainty in the deflected flow rate, and is represented as error bars, since it contributes additional incident kinetic energy not accounted by our idealization.
}
\label{fig:Summary1a}
\end{figure}

In our configuration, the flow reduction occurs for two reasons.
The first is due to resistance to the flow due to the turbine array, which causes more fluid to flow around the turbine array.
As resistance from the actuator disk is the cause of flow reduction in the \LBJ~analysis, we expect the quantitative contribution of this mechanism of flow reduction to be similar.
In the \LBJ~analysis the flow reduction is $Q/Q_0 = \frac{1}{2}\left(1+V/U\right)$, plotted as a dashed black line in Figure \ref{fig:FlowPower}~(b).
Indeed the expected flow reduction for our array at the weak deflection limit, the uppermost red line plotted in Figure \ref{fig:FlowPower}, is close (but not identical) to the \LBJ~line.
Our flow reduction factor nearly follows that of \LBJ~in the weak deflection limit because we use a similar procedure to determine efficiency-- an energy balance on a volume defined by two streamlines shedding vorticity.

The second reason for flow reduction is the development of recirculation regions at the edges of the array.
A detailed characterization of the size of the recirculation regions is presented in Appendix \ref{sec:recirclen}.
As the recirculation regions are small but finite in the limit of weak deflection, the flow reduction factor, $Q/Q_0$ differs only slightly from the \LBJ~case (the difference between the red and black lines).
In the case of strong deflection, the increased turbine array resistance results in larger recirculation regions. 
As a consequence of the reduced effective area of the array, the flow through the array is lower than the \LBJ~case (lines of increasing blue intensity in Figure \ref{fig:FlowPower}~(b)).
Note that the recirculation regions impact the deflected-flow strength as described in \S\ref{subsec:ldta-bd} and efficiency terms which both contribute to $\Ca$ in \eqref{eqn:ca}.

The dependence of the array efficiency $\eta$ on $V/U$ is also plotted in Figure \ref{fig:FlowPower}~(c). 
A maximum $\eta$ for fixed $Q_0/(UL)$ arises as we vary $V/U$ due to a competition between the increasing wake deficit and decreasing flow, which can be defined as
\begin{align}
 \etamax \left(\dfrac{Q_0}{UL}\right)= \max_{V/U} \quad \eta\left(\dfrac{Q_0}{UL}, \dfrac{V}{U} \right),
\end{align}
and plotted in the inset of Figure \ref{fig:FlowPower}~(a).
The array efficiency, $\eta$, and by extension $\etamax$, are bounded between two values and reduce monotonically with the deflected-flow strength.
The highest efficiency corresponds to the weak deflection limit, in which case $\etamax\approx \etaval$ occurs for $V/U=0.38$, close to the values derived from \LBJ~analysis. 
The lowest efficiency corresponds to the strong deflection limit, in which case $\etamax \approx \etavalmin$ occurs for $V/U=0.59$.
The reduction in efficiency in the strong deflection limit may be traced back to a smaller flow reduction factor $Q/Q_0$ for this case, which in turn arises from the increase in size of the recirculation zones due to resistance to the flow offered by the turbines.
In summary, increasing the flow through the array from the weak deflection to the strong deflection limit, decreases the efficiency of the array at extracting the impinging energy from 57\% to 39\%.

A bound on the array power density for the idealized case can be obtained by recognizing that, by definition, $\eta \le \etamax$, and which upon substitution into \eqref{eqn:ca} leads to 
\begin{align}
\Ca \le \dfrac{Q_0}{UL} \etamax. 
\label{eqn:ourlimit}
\end{align}
This expression is plotted in Figure \ref{fig:Summary1a}.

The array power density in \eqref{eqn:ourlimit} has opposing dependences on the flow deflection and turbine efficiency. 
On one hand, the maximum possible efficiency of energy conversion, shown in the inset of Figure \ref{fig:FlowPower}~(a), decreases with the deflected-flow strength.
On the other hand, the stronger the deflected flow, the more kinetic energy impinges on the turbines in the array and is available for extraction.
Array power density is effected more strongly by the increase of impinging kinetic energy due to deflection than the decrease in array efficiency from  $\etaval$  to $\etavalmin$.
Therefore, larger array power densities are possible if the bound vorticity in the deflector-turbine array can be reinforced and greater flow deflected, without influencing the free vorticity.
In other words, an optimal value for the deflected-flow strength does not emerge as part of our analysis; 
the greater the deflected-flow strength, the higher can the array power density be.

\subsection{Power per unit array length}
\label{sec:powerperunitlength}
A consequence of our choice of bound and shed vorticity is that for weak deflection the distribution of extracted power is uniform along the length of the array.
For completeness, we reproduce this result from \cite{Mangan2016a}.
To see this, consider the fluid mechanical energy flux, $\bE = \bu p_0$, where $p_0$ is the stagnation pressure $p + \frac{1}{2} \rho |\bu|^2$. 
The mechanical energy is conserved, and therefore $\bE$ is divergence free, everywhere except on the array.
On the array, the flux jumps discontinuously, by an amount determined by the power extracted by the array.
Furthermore, the Bernoulli principle for inviscid flow implies that the stagnation pressure is one of two values; it is $p_\infty + \frac{1}{2} \rho U^2$ outside the wake, and $p_\infty + \frac{1}{2} \rho V^2$ within the wake.
The stagnation pressure jumps in value only along streamlines that pass through the array but do not recirculate, and this jump is uniformly equal to $\frac{1}{2} \rho (U^2-V^2)$ across the array. 

Therefore, the power converted per unit length along the array is $\wdot_d=0$ if the streamline passing through the point recirculates and $ \wdot_d = \frac{1}{2} \rho (U^2-V^2) \bu\cdot\bn$ otherwise. 
(Here the subscript $d$ denotes a dimensional quantity, which we render dimensionless later.)
The variation of the power density with location arises only from the profile of normal component of the fluid velocity.
Integrating this quantity over the length of the array yields 
\begin{align}
\Wdot = \int_{-L/2}^{L/2} \wdot_d~\diff s = \frac{1}{2} \rho (U^2-V^2) \int_{s_1}^{s_2} \bu\cdot\bn~\diff s = \frac{1}{2} \rho (U^2-V^2) Q,
\end{align}
which is identical to \eqref{eqn:kedeficit}.
We obtain the normal component of velocity through our numerical method described in \S\ref{sec:nummeth}, and determine the profile of power density.  
We use the deflected flow rate $Q_0$, which occurs in the absence of any energy extraction, to non-dimensionalize the power density as
\begin{align}
\wdot = \dfrac{\wdot_d}{{\frac{1}{2L} \rho U^2 Q_0}}.
\end{align}

\begin{figure}
\centerline{\includegraphics{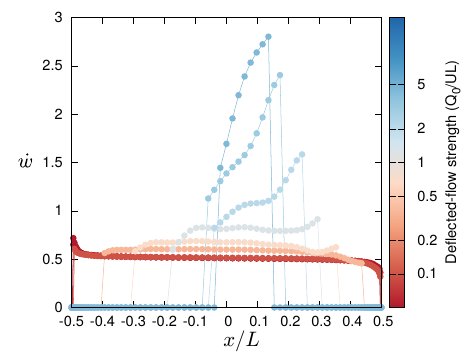}}
\caption{(Colour online) The power extraction per unit length as a function of location along the array. Colour corresponds to the value of deflected-flow strength.}
\label{fig:TurbineDensity}
\end{figure}

The distribution of $\wdot$ with location along the array corresponding to the optimal energy extraction as a function of given deflected-flow strength is plotted in Figure~\ref{fig:TurbineDensity}.
In the limit of small deflection, $\bu\cdot\bn \approx $ constant = $U\tan\aoa$, leading to an uniform distribution of power extraction. 
The uniform value agrees with the optimal array efficiency $\etamax$ for that deflected-flow strength.
However, for stronger deflected flow strength, $\wdot$ is not uniform along the array -- we have no further insight on the shape of these curves.


\begin{figure}
\centerline{\includegraphics[width=0.8\textwidth]{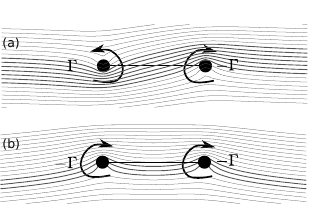}}
\caption{Illustration of the significance of anti-symmetric bound vorticity distribution. 
(a) Streamlines due to bound vorticity comprised of uniform flow superposed with two equal and opposite point vortices (filled circles with curved arrows) located at the ends of the array (thick solid line) in a uniform freestream. 
Bold streamlines constitute a streamtube that passes between the two vortices.
(b) Same as (a) but due to bound vorticity comprised of two equal point vortices.
Bold streamlines constitute the streamtube that passes through the two vortices.
}
\label{fig:Conclusion}
\end{figure}

\subsection{Comparison with example deflector-turbine arrays}
\label{subsec:comparison}
The linear deflector-turbine array is derived as an idealization of more general deflector-turbine arrays confined to narrow rectangles aligned with the freestream, e.g. the ones in \S\ref{subsec:deflectors}-\ref{subsec:vawt}.
The vorticity and velocity fields established in the linear deflector-turbine array are a simplified representation of those established in both these examples.
Consequently, we expect the power extraction performance of both the example arrays to be approximated by the power extraction performance of the linear deflector-turbine array.
While the maximum power extracted in both the examples increases with the flow deflection (upto stages described respectively in \S\ref{subsec:deflectors}-\ref{subsec:vawt}), we ask next whether these qualitative correspondences rise to the level of quantitative agreement.
To answer this question, we compare the relation between the deflected-flow rate $Q_0$ and the computed maximum value of $\Wdot$, both non-dimensionalized to the deflected-flow rate $Q_0/(UL)$ and $\Ca$, for the three systems.

The results are plotted in Figure \ref{fig:Summary1a}.
For both example arrays, the values of $Q_0/(UL)$ are calculated from the case when the turbines are turned off.
For \S\ref{subsec:deflectors}, the value of the $Q_0$ depends on the angle of attack, $\defaoa$, and is calculated as -$\int_{-2}^2 v~\diff x$.
The range of integration spans over the array and the negative sign accounts for the flow being deflected along the negative $y$ direction.
However, there is ambiguity about the treatment of the flow $\Delta Q_0 = \int_{-0.2}^{0.2} u~\diff y$ directly incident on the frontal area; we treat this as our uncertainty in $Q_0$ because this additional flow imparts additional kinetic energy flux on the turbines but is neglected by our idealization.
For each $\defaoa$, the turbine induction factor is varied by changing $\perm$, and the maximum extracted power $\Wdot$ is determined.
Both $Q_0$ and $\Wdot$ are then non-dimensionalized to calculate the values of $Q_0/(UL)$ and $\Ca$, using $L=4$ as the length of the array. 
For the example array in \S\ref{subsec:vawt}, $Q_0$ depends on the value of $\beta$ with $\perm =0$ and is calculated as $\int_{-19}^{1} v~\diff x$, where the range of integration accounts for the extent of the array.
In addition, the flow incident on the frontal area $\Delta Q_0 = \int_{-1}^{1} u~\diff y$ is considered as the uncertainty in $Q_0$.
The maximum $\Wdot$ is determined by varying $\perm$ whilst maintaining a constant value of $\beta$.
The values of $Q_0/(UL)$ and $\Ca$ are determined by non-dimensionalizing $Q_0$ and $\Wdot$, where $L=20$ is the length of the array in this case.
Since our idealization ignores the frontal area of the array, only cases where the deflected flow is greater than the flow impinging on the frontal area of the array are included.
The computed values for $\Ca$ increase monotonically with increasing $Q_0/(UL)$ for both the arrays, and then decrease for the last data point.
For the array in \S \ref{subsec:deflectors}, the decrease occurs because the boundary layer separates from the airfoils, while for that in \S \ref{subsec:vawt}, it is because the turbines do not intercept all the deflected streamlines.
The remaining values agree well with those resulting from the idealized array, despite the differences in geometry, the turbine representations and the deflection mechanism.


To understand this agreement, consider how a slender stream tube of fluid passing through the array accumulates energy deficit.
The rate of deficit accumulation depends on the profile of resistive force the fluid encounters, which in turn depends on the distribution of turbines and deflectors along its path.
However, note that according to \cite{Newman1986}, the optimal array performance barely increases from 16/27 to 2/3 when the number of turbines an isolated stream tube encounters increases from 1 to $\infty$.
These two observations may be qualitatively combined to infer that the energy extraction from a stream tube is insensitive to the number or distribution of turbines it encounters so long as a near-optimal deficit accumulates on it.
The linear deflector-turbine array is successful because it idealizes away the distribution of turbines a streamline encounters in its path and replaces it with a near-optimal energy deficit as it crosses the array.
Our inviscid flow calculation also accounts for the interaction between neighbouring stream tubes and the influence of flow deflection as they pass through the array, causing the efficiency to reduce from 57\% to 39\%, but no further.
The idealized result \eqref{eqn:ourlimit} does not constitute a strict upper limit on the performance of the non-ideal cases, the difference depending on the degree to which the underlying assumption are relaxed.
It is the reason why some portion of the data error bars in Figure \ref{fig:Summary1a} exceed \eqref{eqn:ourlimit}.
The value of the idealized model is (i) in identifying the performance parameters, in this case $Q_0/(UL)$ and $\Ca$, (ii) in rationalizing the nature of the maximum power extraction, i.e. $\Ca < \eta Q_0/(UL)$, and (iii) in determining the approximate value for $\eta$.
Based on the agreement with the computational arrays in \S\ref{subsec:deflectors}-\ref{subsec:vawt}, we believe that the linear deflector-turbine array provides an approximate but insightful parametrization for the maximum power extraction.

\subsection{Asymmetric distribution of bound vorticity}
\label{subsec:mechanistic}
The key feature of the bound vorticity profile given by \eqref{eqn:boundvorticity} is its anti-symmetry about the array center; i.e. the bound vorticity in the upstream half of the array (i.e. $x<0$) has the opposite sign compared to the downstream half (i.e. $x>0$).
Figure~\ref{fig:Conclusion} illustrates the significance of this feature using two point vortices. 
Figure~\ref{fig:Conclusion}~(a) shows the superposition of a uniform flow and two equal and opposite point vortices.
A band of streamlines is deflected between the point vortices.
Consequently, the streamlines that pass through the first black circle (labeled $\Gamma$) flow above and avoid the second black circle (labeled $-\Gamma$).
In contrast, Figure~\ref{fig:Conclusion}~(b) shows the streamlines for two identical point vortices.
In this case, the streamlines are curved, but not deflected around the downstream vortex.
The streamlines passing through the first black circle (labeled $-\Gamma$) are directly incident on the second black circle (also labeled $-\Gamma$).
For more visualizations of the flow resulting from symmetric and anti-symmetric arrangements of four vortices see \cite*{Kirshen2016}.

In general, the two halves of an anti-symmetric profile of vorticity induce a velocity of the same sign normal to the array.
This induced velocity deflects the flow.
More precisely, the deflected flow due to a linear deflector-turbine array may be written in terms of the bound vorticity distribution as
\begin{align}
 Q_0 = \int_{-L/2}^{L/2} \int_{-L/2}^{L/2} \dfrac{1}{2\pi} \dfrac{\vortb(s)}{x-s}~\diff s~\diff x 
 = \int_{-L/2}^{L/2} \dfrac{\vortb(s)}{2\pi} \log \left( \dfrac{L/2-s}{L/2+s} \right)~\diff s.
 \label{eqn:Q0onVortB}
\end{align}
Here we have ignored the influence of the re-circulation near the array ends assuming it to be small, as is the case in the weak deflection limit (see Appendix \ref{sec:recirclen}).
Next we decompose the bound vorticity into its symmetric and anti-symmetric part $\vortb(s) = \omega_{b,s}(s) + \omega_{b,a}(s)$, where $\omega_{b,s}(s) = \tfrac{1}{2} ( \vortb(s) + \vortb(-s) )$, and $\omega_{b,a}(s) = \tfrac{1}{2} ( \vortb(s) - \vortb(-s) )$.
Due to the anti-symmetry of the logarithmic term's dependence on $s$, the contribution of $\omega_{b,s}(s)$ to $Q_0$ is zero, and the integral \eqref{eqn:Q0onVortB} derives its value solely from $\omega_{b,a}(s)$.
In this manner, the degree of asymmetry in the bound vorticity profile is generically linked to flow deflection.
Based on this intuition, we arranged the deflectors in \S\ref{subsec:deflectors}-\ref{subsec:vawt} to induce an asymmetric distribution of bound vorticity.

A similar rationale underlies the array design by \cite*{Corten2007}, who suggested arranging non-uniform yaw and pitch of axial flow wind turbines in an array, and non-uniform angular speed distribution on cross-flow turbines.
They also considered embedding deflector surfaces in the turbine arrays.
These strategies suggest an asymmetric variation of the controlled quantities (yaw or pitch of axial flow turbines and angular speed of cross flow turbines) within the array.
If these variations cause an asymmetric bound vorticity profile then they will cause flow deflection as described in this article.
However, key differences render our understanding of the strategies in \cite{Corten2007} to be suggestive at best.
These differences include the three-dimensional nature of their strategies versus the two-dimensional nature of our analysis.
Furthermore, it is not clear whether the magnitude of bound vorticity induced by changing yaw and pitch of axial flow turbines or the angular speed of cross-flow turbines is adequately strong, and requires additional investigation.

\section{Application to hydrokinetic power}
\label{sec:hydrokinetic}
\begin{figure}
\centerline{\includegraphics[width=0.98\textwidth]{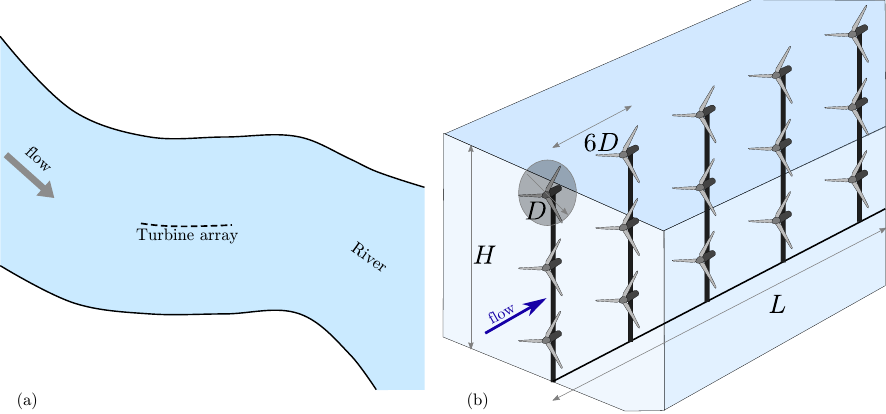}}
\caption{Schematics showing a hypothetical application for hydrokinetic energy proposed by \cite{Mandre2016a}. 
(a) Plan view of a narrow constriction in a river is a hotspot for hydrokinetic power. 
An deflector-turbine array is constrained to a narrow strip shown by the dashed line, aligned along the centerline and spanning the depth of a river.
(b) A conventional array of axial-flow turbines of diameter $D$ forming a vertical fence along the river centerline. 
The water depth is $H$, the array length is $L$, and the turbines are spaced $6D$ apart, implying the number of turbines to be $LH/(6D^2)$.
}
\label{fig:RiverineSchematic}
\end{figure}

Our analysis is motivated by pragmatic constraints in riverine and tidal hydrokinetic turbine arrays.
One way to accommodate turbine arrays in prime, hydrokinetic-energy hotspots is to limit their spatial profile across the river, such as the one shown in Figure~\ref{fig:RiverineSchematic}~(a). 
In this spirit, \cite*{Mandre2016a} proposed a hypothetical array consisting of crossflow turbines and oscillating hydrofoils, installed in a narrow strip along the centerline of a river shown in Figure \ref{fig:RiverineSchematic}~(a).
Here we consider a vertical fence of axial-flow turbines spanning the depth of the water column along the river's centerline, as shown in Figure~\ref{fig:RiverineSchematic}~(b).
To ascertain the applicability of our analysis, we derive a hueristic condition for the neglect of the energy flux incident on the array due to turbulence and the background pressure gradient.

The array is composed of axial-flow turbines of diameter $D$ in a dense $m \times n$ rectangular grid, where $m=H/D$ and $n=L/(6D)$, installed throughout the water column of depth $H$ and forming an array of length $L$.
(The factor of 6 in the expression for $n$ is explained later.)
Without turbulence and flow deflection, the analysis by \cite{Newman1986} predicts that at most fraction $8n(n+1)/(3(2n+1)^2)$ of the incident kinetic energy $\rho U^3 m\pi D^2/8$ could be extracted.
As discussed earlier, installing more rows of turbines only provides marginal returns according to \cite{Newman1986}, and increasing the vertical spacing reduces $m$ and consequently the array power.
However, turbulent processes replenish the wake of upstream turbines and increase the incident power on the turbines.
While a complete treatment of the influence of turbulence is outside the scope of this article, the condition we derive uses an estimate of this turbulent kinetic energy flux and compares it with the deflected kinetic energy flux.

We estimate this role of turbulence mixing in a manner presented by \cite{Mackay2008} as follows.
We assume that axial-flow turbines are spaced six diameters apart.
(This explains the factor of $6$ in the expression for $n$.)
Experiments by \cite{Jeffcoate2016} report an experimental investigation of two in-line axial-flow hydrokinetic turbines separated by six diameters. 
The downstream turbine generated at most 40\% of the maximum power generated by the upstream turbine, implying that the wake deficit had not recovered.
However, we assume that the turbine wakes recover completely before impinging on downstream turbines. 
We assume implicitly that turbines spaced closer than this will have a lower power density, and doing so allows us to derive an empirical but conservative over-estimate of the influence of turbulence. 
Therefore, for the case shown in Figure~\ref{fig:RiverineSchematic}~(b), the array power is bounded by 
\begin{align}
\Wdot_\text{3D,turbulent} = \eta_\text{\LBJ} \dfrac{LH}{6D^2} \times \dfrac{\rho U^3 \pi D^2}{8} = \eta_\text{\LBJ} \dfrac{LH \rho U^3 \pi}{48},
\end{align}
where $\eta_\text{BJ}=16/27$ is the Betz-Joukowsky efficiency.
Note that this expression is independent of the turbine diameters because the number of turbines in the array is inversely proportional to their individual frontal area.

To investigate the role of flow deflection we assume that, despite the three-dimensional internal structure of the array and a non-uniform depth-wise profile of the channel flow, our two-dimensional analysis could provide a useful approximation.
(Our approximation is similar in motivation to the two-dimensional approximation to channel flows by shallow water equations.)
If  the role of turbulence is negligible and the array length is much smaller than the channel width, then our analysis provides guidelines for the appropriate bound vorticity to deflect the flow through the array.
We imagine flow deflection caused by airfoils in a manner similar to the case of \S\ref{subsec:deflectors}.
The number of turbines is assumed to be large enough so that all the deflected flow is intercepted by one or more turbines. 
If a flow $Q_0H$ (in three dimensions) is deflected through the array, then according to our results an amount 
\begin{align}
\Wdot_{3D,deflected} = \eta_\text{non-ideal} \etamax \dfrac{\rho U^2 Q_0 H}{2},
\end{align}
where $\eta_\text{non-ideal}$ is a fraction between 0 and 1 accounting for the departure from ideal behavior, could be extracted.
The ratio of the two is 
\begin{align}
\dfrac{\Wdot_{3D,deflected}}{\Wdot_\text{3D,turbulent}} = 24 \eta_\text{non-ideal} \dfrac{\etamax}{\eta_\text{\LBJ}} \times \dfrac{Q_0}{UL}.
\end{align}
The effect of flow deflection is expected to dominate when the right hand side is much greater than unity, or
\begin{align}
\dfrac{Q_0}{UL} \gg \dfrac{1}{24} \dfrac{\eta_\text{\LBJ} }{\eta_\text{non-ideal} \etamax} \ge \dfrac{0.063}{\eta_\text{non-ideal} }.
\label{eqn:neglectturbulence}
\end{align}
If we assume $\eta_\text{non-ideal} =0.8$, the right hand side is approximately 0.08.
$Q_0/(UL)$ greater than 0.08 are feasible using airfoils as deflectors with $\defaoa\ge 6^\circ$, as shown in Figure~\ref{fig:Summary1a} for the example array in \S\ref{subsec:deflectors}.
As noted above the influence of turbulence in deriving \eqref{eqn:neglectturbulence} is grossly overestimated.

Next we investigate the influence of the background pressure gradient, in this case driven hydrostatically by the slope, $S$, of the river.
Over a length $L$, the hydrostatic pressure drop is $\rho g SL$.
This value is negligible compared to the dynamic pressure, which scales as $\rho U^2$, if
\begin{align}
L \ll \dfrac{U^2}{gS}.
\label{eqn:neglectslope}
\end{align}
For typical values, $U=2$ m/s and $S=10^{-3}$, the hydrostatic pressure gradient is negligible is $L \ll 400$ m.

\section{Discussion}
\label{sec:discussion}
Through steady inviscid flow analysis, we have developed definitions of deflectors and turbines based on the type of vorticity they introduce in the flow.
In this framework deflectors in a compact region cause bound vorticity and modify the flow close to the array, but due to the decay of induced velocity with distance, fail to generate a wake deficit.
Turbines shed free vorticity, and thereby result in wake reduction far downstream.
We further show that for linear arrays, the fraction of deflected kinetic energy extracted by the turbines lies between 39 to 57\% depending on how strongly the flow is deflected. 
The deflection increases the incident kinetic energy, outweighing the reduction in efficiency, and the design of arrays could, therefore, benefit from enhancing flow deflection.
The general arrangement of deflectors that enhance flow deflection introduce an asymmetric distribution of bound vorticity.
Our analysis is limited to two-dimensional flows, which arise in specific applications discussed in \S\ref{sec:hydrokinetic}.

Due to their basis in vorticity dynamics, our results are agnostic to the specific deflecting and energy converting mechanisms employed in the array.
The most obvious candidates for manipulating the flow is an array of static airfoils interspersed with the turbines. 
A host of other conceivable techniques have been discussed in the literature for three-dimensional flows but would be equally applicable in two. 
These include turbines oriented at an angle to the flow \citep{Corten2007,Wagenaar2012,Fleming2014,Churchfield2015}, cross-flow turbines \citep{Dabiri2011a,Kinzel2012a,Mandre2016a}, or foils oscillating asymmetrically about their mean positions, to vector its wake at an angle to the free stream.
Static airfoils would be limited to regimes where the boundary layer on them remains attached, as observed for the example array in \S\ref{subsec:deflectors}.
Another limit on the performance enhancement arises when turbines are spaced too far apart and they fail to intercept all the deflected flow, as observed for the example array in \S\ref{subsec:vawt}.
The details of the other deflection mechanisms will certainly result in additional limits on the deflected-flow strength.
In any case, our analysis shows that a successful array is composed of two essential components: (i) bound vorticity to deflect the flow and (ii) free vorticity to intercept the deflected kinetic energy.

Our framework also presents approaches to simplify models of cross-flow turbines presented in the literature \citep*{Islam2008,Whittlesey2010,Araya2014}.
Note that the definitions developed in \S\ref{sec:defdefturb} do not differentiate between a single turbine and an array, and therefore also apply to simplified models of two-dimensional turbines.
According to the analysis in \S\ref{sec:defdefturb}, cross-flow turbines operating in an hypothetical inviscid fluid would also influence the flow by inducing bound vorticity and shedding free vorticity.
The bound vorticity that exists by virtue of the rotation of the cross-flow turbine may be a significant influence on the surrounding flow even when operating in a viscous fluid.
Note that the body force in \eqref{eqn:vawtbf} presents a prescriptive method for imposing (possibly non-uniform) bound vorticity profiles.
Models of the type presented in \S\ref{subsec:vawt} may be developed, perhaps with the non-uniformly distributed body force, to better approximate the influence of a cross-flow turbine on the flow.

We neglect turbulence, the predominant mechanism for wake replenishment, because our premise is to investigate wake deflection.
Our analysis is most directly applicable to tight arrays, in which turbulent mixing is too slow to influence the wake and flow deflection dominates energy transport. 
The quantitative validity of our analysis relative to the role of turbulence in the general case depends on the details of the types of turbines, array geometry, and the active turbulent processes.
We derived heuristic conditions for the neglect of turbulence for the case of the hydrokinetic array shown in Figure~\ref{fig:RiverineSchematic}~(b).
Detailed analysis, outside the scope of this article, needs to be conducted to characterize these effects in general.

In our analysis, we have neglected external boundaries on the fluid domain, in order to focus on the essential principles.
For an array length comparable with the channel width, the presence of boundaries will influence the flow and the performance of the array.
However, for arrays shorter than channel width, the influence of the boundaries can be neglected, and our analysis is applicable.
The omission of boundaries is not a fundamental limitation of our approach or our analysis.
Boundaries can be incorporated using the method of images or boundary integral methods for potential flow, resulting in improved bounds and better constraints on array design.

The extension of our framework to three-dimensional flows requires incorporation of more general vorticity dynamics.
Just as lifting line theory for three-dimensional wings is based on the two-dimensional Kutta-Joukowsky relation between bound vorticity and airfoil lift, our results in two dimensions provide the foundation for such an extension to three dimensions.
Furthermore, unsteadiness and turbulence, the role of fluid viscosity, presence of shear in the freestream, the presence of ground and other boundaries, and the discrete nature of the array will reduce the performance below the limits we present here.
We have also not accounted for gravity, the free interface, and bed friction for the hydrokinetic application.
These effects could be examined more carefully by embedding the deflector-turbine array in a larger computational fluid dynamics model and through controlled experiments.
However, the semi-analytic expressions calculated here provide an avenue for differentiating and quantifying these more complicated effects relative to the ideal case. 

\section{Conclusion}
We presented systematic flow manipulation as a strategy to deflect untapped kinetic energy towards a deflector-turbine array.
We use the framework of inviscid fluid dynamics to decompose the effects of array on fluid flow into deflection, captured by bound vorticity, and energy extraction, captured by shed vorticity. 
The results calculated here approximate upper bounds on performance that are agnostic to the specific array technology.
The large gain in kinetic energy deflected onto the array (deflected-flow strength) outweighs the small decrease in array's efficiency at extracting that energy (from \etapercent\% - \etapercentmin\%) due to increased recirculation.
We expect the essential ingredients of this analysis to inform more complicated methods of flow deflection and three-dimensional situations.

Future investigations building on this analysis will help evaluate the technical and economic feasibility of this approach.



\section*{Acknowledgements}
We acknowledge support from the DOE ARPA-E grant DE-AR0000318 for this work. 




\appendix
\section{The role of turbulence models in \S 2}
\label{sec:turbmodel}

\begin{figure}
\includegraphics[width=0.98\textwidth]{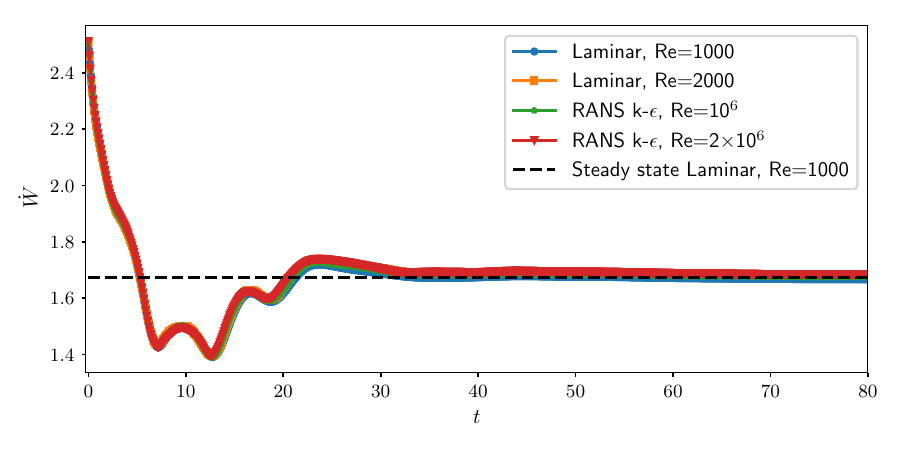}
\caption{Transient power production by the array in \S\ref{subsec:vawt} using two different turbulence models for two Reynolds numbers each. Case $\beta=0.4$, $a=0.2$.}
\label{fig:Turb_TransPower}
\end{figure}

\begin{figure}
\centerline{\includegraphics[width=0.75\textwidth]{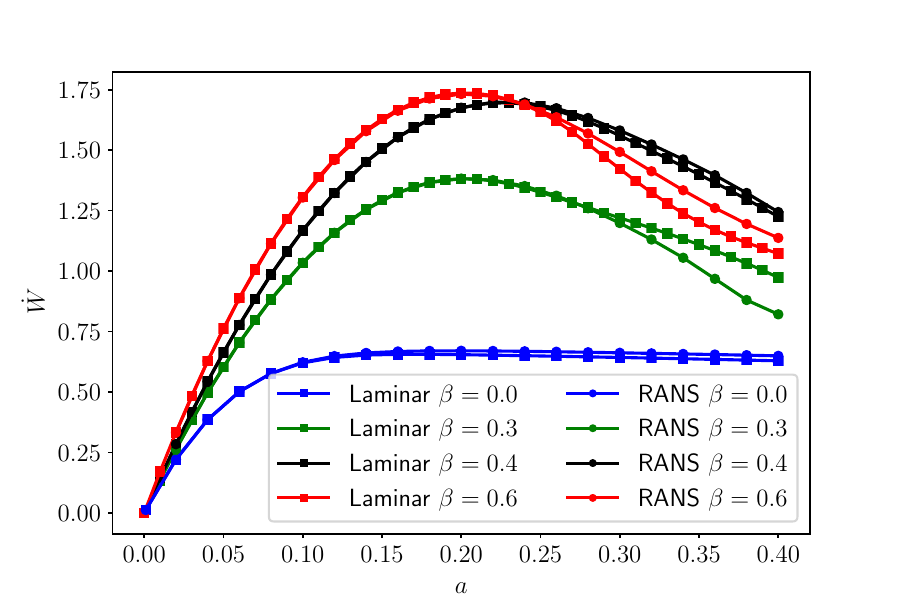}}
\caption{Comparison of the performance of the array is \S 2.2 between laminar flow and RANS models.}
\label{fig:ransvslaminar}
\end{figure}

Within the framework of Reynolds-averaged Navier-Stokes (RANS), the turbulent flow variables are decomposed into a temporal mean and fluctuations. 
The mean velocity components $u_i$ satisfy
\begin{align}
\rho \left( \dfrac{\partial u_i}{\partial t} + u_j \dfrac{\partial u_i}{\partial x_j} \right) + \dfrac{\partial p}{\partial x_i} = f_i + \dfrac{\partial\tau_{ij}}{\partial x_j}, \text{ and } \dfrac{\partial u_j}{\partial x_j} = 0,
\end{align}
where $\tau_{ij}$ are components of the stress, including the Reynolds stress, and indicial notation is used with $i,j$ ranging from 1 to 2.
The $k-\epsilon$ model parametrizes the stress as
\begin{align}
 \tau_{ij} = (\mu + \mu_T) \left( \dfrac{\partial u_i}{\partial x_j} + \dfrac{\partial u_j}{\partial x_i} \right), \quad \text{where} \quad \mu_T = \dfrac{\rho C_\mu k^2}{\epsilon},
\end{align}
$k$ being the turbulent kinetic energy field, $\epsilon$ the turbulent dissipation field, which are governed by their own partial differential equations, $\mu$ is the fluid viscosity and $C_\mu = 0.09$ is a dimensionless constant of proportionality \citep{COMSOLCFD2016}.
The Reynolds number is defined as $\text{Re} = \rho V R/\mu$, where $R$ is the turbine radius and $V$ the freestream speed.
In this model, the turbulent viscosity is a spatio-temporally varying field.
A constant eddy viscosity model for turbulence can be considered mathematically equivalent to the special case where $\mu_T$ is not a spatio-temporally varying field but a constant.
To simulate this case, we use COMSOL's laminar flow solver and set $\mu=\mu_T=$ constant.
Because of our use of the laminar flow solver, we label this case as ``Laminar'' in Figures \ref{fig:Turb_TransPower} and \ref{fig:ransvslaminar}.
Indeed, there is no mathematical difference between a laminar flow and a turbulent flow with a constant eddy viscosity, provided the molecular viscosity and the eddy viscosities agree.

We first present results for a representative parameter set ($a=0.2$, $\beta=0.4$) using the two models with two different Reynolds numbers for each.
Two of the four cases correspond to a constant viscosity with Re=1000 and 2000, respectively, and the remaining two to a RANS $k-\epsilon$ model with Re=$10^6$ and $2\times 10^6$, respectively.
In all four cases, the computation is started from a uniform flow and the body force according to equation \eqref{eqn:vawtbf} is turned on at $t=0$, and the time-resolved solution of the governing equations monitored.
The inlet turbulence is assumed to be the minimum necessary to ensure numerical stability.
The transient evolution of the flow for all the four cases is visualized in the Supplementary Material - Movie 3 and the power extracted by the array is presented in Figure \ref{fig:Turb_TransPower}.
The power transient differs between the four cases by less than 1\%.
All four power transients reach a steady state, which is well-approximated by the steady state solution of ``Laminar Re=1000'' case.

\begin{figure}
\includegraphics[width=0.98\textwidth]{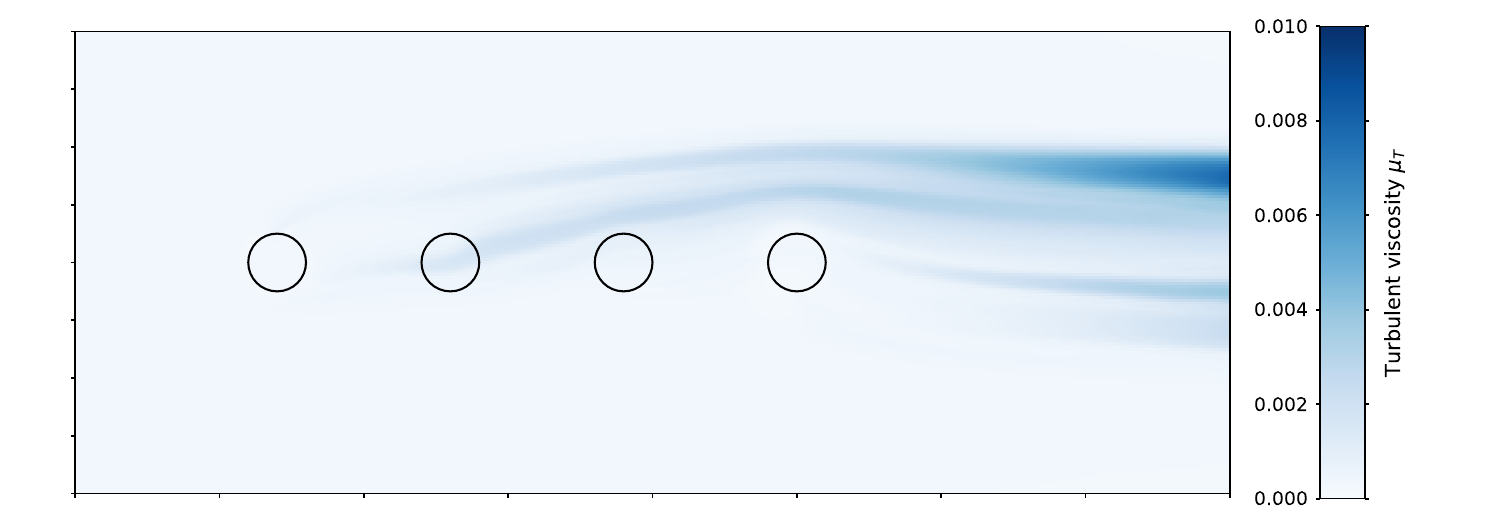}
\caption{The steady state distribution of the turbulent viscosity $\mu_T$ in the RANS Re=10$^6$ case in Figure \ref{fig:Turb_TransPower}.}
\label{fig:muT}
\end{figure}

The reason for this agreement may be understood by examining the turbulent viscosity ($\mu_T$) field, which constitutes the difference between laminar and the RANS model.
Figure \ref{fig:muT} shows the steady state $\mu_T$, which is approached as $t\to \infty$. 
The turbulent viscosity is strongest downstream of the deflector-turbine array along the shear layer generated by it.
According to the $k-\epsilon$ model, the turbulent viscosity is proportional to the turbulent kinetic energy squared and is highest where the turbulent kinetic energy is highest.
And the turbulent kinetic energy is fed by the mean shear, which is strongest in the shear layers.
Therefore, it is reasonable to expect the turbulent viscosity to be highest downstream of the array along the shear layers.

Despite the strong shears generated in our solutions, the turbulent viscosity always remains weak in the region close to the array.
The turbulent viscosity, non-dimensionalized by $\rho V R$,  is everywhere less than $7 \times 10^{-3}$, indicating the turbulent stresses are about 1\% or less of the inertial forces in momentum balance.
Therefore, the turbulent stresses only make higher-order contributions to the performance of the deflector-turbine array.
The leading order is determined by inertial forces, which are modeled as accurately by a laminar flow model as they are by a turbulent flow model.
In this manner, we deduce that turbulence is a weak influence in our computational solutions.

To ensure that this argument does not depend on the specific parameters we chose in Figures \ref{fig:Turb_TransPower} and \ref{fig:muT}, we repeated the analysis for parameters spanning the range presented in the manuscript.
Figure \ref{fig:ransvslaminar} shows the power generated by the array for $0\le \beta \le 0.6$ and $0\le a \le 0.4$ using both a laminar flow model and a RANS $k-\epsilon$ model.
The difference is greatest when the deflection is weak ($\beta$ around 0.3) and the drag is large $a>0.3$.
Regardless, the difference is everywhere $<25\%$, and for values of $a$ less than or around the optimal is $<1\%$. 

A similar analysis for the influence of turbulence in the solutions presented in \S\ref{subsec:deflectors} reveals that the strength of turbulent viscosity is comparable to inertial forces only in the airfoil boundary layers where the shear is strongest.
This influence of turbulence maintains an attached boundary layer upto $\defaoa=12^\circ$, where the laminar flow model with Re=1000 results in the boundary layer separating around 6-8$^\circ$.
In comparison to the inertial forces, the turbulent eddy viscosity term in momentum balance remains $<1\%$ outside the airfoil boundary layer.
Therefore, the difference between the solutions computed using RANS and the laminar flow models are negligible, so long as the boundary layer on the airfoils remains attached.

\section{Solution using complex variables}
\label{sec:complexsol}
The function
\begin{align}
 w(z) = U z + i U \tan\aoa \left( z - \sqrt{z^2-L^2/4} \right),
\end{align}
of the complex variable $z=x+iy$ serves as the complex potential for the flow we seek. 
The stream function $\psi$ is the imaginary part of $w$. 
The velocity field arising from this flow is
\begin{align}
\overline{\dfrac{\diff w}{\diff z}} = U + i U\tan \aoa \left( 1 - \dfrac{\zbar}{\sqrt{\zbar^2-L^2/4}} \right),
\end{align}
where the overbar denotes complex conjugation.
The flow velocity just above and below the array is
\begin{align}
 \overline{\dfrac{\diff w}{\diff z}}\left( z = x + i 0^{\pm} \right)  = U + i U \tan\aoa \pm U \tan\aoa \dfrac{x}{ \sqrt{L^2/4 - x^2} }.
\end{align}
This expression implies the mean velocity makes an angle $\aoa$ with the freestream, and the tangential speed jumps discontinuously by $2 Ux \tan\aoa /\sqrt{L^2/4-x^2}$.
This jump corresponds to the negative of the bound vortex sheet strength shown in \eqref{eqn:boundvorticity} in the paper.
The uniqueness of the solution of Laplace equation implies that the solution we present here is the only possible solution.

An example of the flow field derived using this technique is shown in Figure~\ref{fig:Summary1}~(a) of the article. 
The flow develops two recirculation regions centered around the end points of the array $x=\pm L/2$, and the fluid from far upstream is entrained into the array between these regions.
The streamline that bounds the recirculation region forms the separatrix between the flow from the far upstream that does or does not pass through the array.
This separatrix necessarily contains a stagnation point of the flow, the point where the flow velocity is zero.

The condition that the extreme streamlines passing through the array also pass through a stagnation point may be used to determine flow through the turbine array.
The location of the stagnation point, $z_0$ may be determined by using
\begin{align}
 \overline{\dfrac{\diff w}{\diff z}} (z=z_0) = U + i U\tan \aoa \left( 1 - \dfrac{\zobar}{\sqrt{\zobar^2-L^2/4}} \right) \equiv 0.
\end{align}
This condition leads to two stagnation points
\begin{align}
 \dfrac{z_0}{L} = \pm \dfrac{1}{2} \dfrac{1-i\cot\aoa }{\sqrt{-i\cot\aoa (2-i\cot\aoa)}}.
\end{align}
and the stream function value at those points
\begin{align}
 \dfrac{\psi(z=z_0)}{UL} = \pm \dfrac{1}{2} \Re \left(\sqrt{-1-2i\tan\aoa} \right).
\end{align}
The difference between the two values of stream function, after representing it in real variables, is the flow rate through the array given by \eqref{eqn:zeroflow} in the paper.

\section{Length of the recirculation region}
\label{sec:recirclen}
\begin{figure*}
\centerline{\includegraphics[width=\textwidth]{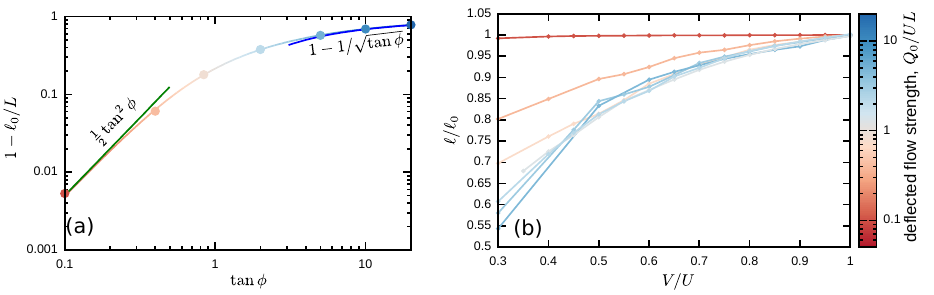}}
\caption{(Colour online)
Characterization of the length $\ell_0$ without energy conversion and $\ell$ with energy conversion of the region on the turbine between the two recirculation regions. 
(a) The length of array occupied by the recirculation region $L-\ell_0$ scaled by the array length $L$ is plotted as a function of the deflector slope, when the flow is merely deflected but no energy is extracted. 
In the weak deflection limit, the asymptote $\ell_0 \approx L(1- \tan^2\aoa/2)$ is also shown. 
In the strong deflection limit, $\ell_0/L \approx 1/\sqrt{\tan \aoa}$, and the recirculation region occupies almost the whole length of the array.
(b) The reduction in the length $\ell$ compared to its value, $\ell_0$,  without energy extraction.
As the wake velocity reduces below the free stream value, the size of the recirculation region grows, and the length of the region in between shrinks. 
This reduction is a small but finite in the limit of weak deflection, and the reduction is prominent in the limit of strong deflection.
}
\label{fig:Recirc}
\end{figure*}

In this section, we provide more details about dependence of the length of the recirculation region on the flow parameters. 
Results from our numerical solutions characterizing this length is presented in figure \ref{fig:Recirc}.
The quantity of interest is the length, $\ell_0$ without energy extraction and $\ell$ with energy extraction, of the segment on the array between the two recirculation regions. 
Figure \ref{fig:Recirc}~(a) shows the dependence of the length of the array occupied by the recirculation region scaled by the array length.

The length $\ell_0$ may be determined analytically by using \eqref{eqn:zeroflow} in the article and the relation
\begin{align}
 U \tan \aoa \ell_0 = Q_0.
\end{align}
Based on this relation, the two asymptotic limits may be deduced.
In the limit of weak deflection, $\ell_0/L \approx 1-\tan^2\aoa/2$, and in the limit of strong deflection $\ell_0/L \approx 1/\sqrt{\tan\aoa}$. 
These asymptotes are also shown in Figure \ref{fig:Recirc}~(a).

For wake speed less than the free stream, the recirculation region grows, and consequently $\ell$ is smaller than $\ell_0$.
This dependence of $\ell$ on $V/U$ is shown in Figure \ref{fig:Recirc}~(b).
In the weak deflection limit, the recirculation region grows to a small but finite size compared to the array length.
In the strong defection limit, the recirculation region grows and $\ell$ decreases prominently below $\ell_0$. 
This growth in the size of the recirculation regions underlies the quantitative difference between \LBJ ~and our results.

\section{Biot-Savart line integral over a segment}
\label{sec:biotsavartsegment}

\begin{figure}
\centerline{\includegraphics{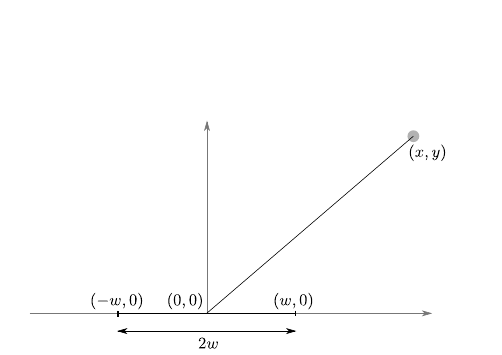}}
\caption{Local coordinate system for evaluating the line integral \eqref{eqn:biotsavartpsi} over a linear segment.}
\label{fig:LineIntegral}
\end{figure}

For evaluating the integral over one segment with unit vortex sheet strength, we define a local coordinate system with origin at the center of the segment and the $x$ axis aligned along its length, so that the segment occupies $-w<x<w$ as shown in Figure \ref{fig:LineIntegral}. 
The contribution to the integral from this segment for any target point $(x,y)$ in the local coordinate system may be written as
\begin{align}
 I(x, y; w) = \int_{-w}^w ~ \dfrac{ \log ((x-s)^2+y^2)}{4\pi} \diff s,
\end{align}
where $s$ is the arc-length coordinate along the segment. 
This integral may be evaluated to be
\begin{align}
\begin{split}
 I = \dfrac{1}{4\pi} \left[ (x+w) \log ((x+w)^2 + y^2) - (x-w) \log((x-w)^2+y^2) - 4w \right. \\
 \left. + 2 y \left( \text{arctan} \dfrac{x+w}{y} - \text{arctan} \dfrac{x-w}{y}  \right) \right].
\end{split}
\end{align}

\bibliographystyle{jfm}
\bibliography{Beyond_Betz}

\begin{thebibliography}{41}
\expandafter\ifx\csname natexlab\endcsname\relax\def\natexlab#1{#1}\fi
\def\au#1{#1} \def\ed#1{#1} \def\yr#1{#1}\def\at#1{#1}\def\jt#1{\textit{#1}}
  \def\bt#1{#1}\def\bvol#1{\textbf{#1}} \def\vol#1{#1} \def\pg#1{#1}
  \def\publ#1{#1}\def\arxiv#1{#1}\def\org#1{#1}\def\st#1{\textit{#1}}

\bibitem[Araya {\em et~al.\/}(2014)Araya, Craig, Kinzel \& Dabiri]{Araya2014}
{\sc \au{Araya, D.~B.}, \au{Craig, A.~E.}, \au{Kinzel, M.} \& \au{Dabiri,
  J.~O.}} \yr{2014}  \at{{Low-order modeling of wind farm aerodynamics using
  leaky Rankine bodies}}.  \jt{Journal of renewable and sustainable energy}
  \bvol{6}~(6),  \pg{063118}.

\bibitem[Batchelor(2000)]{Batchelor2000}
{\sc \au{Batchelor, G.~K.}} \yr{2000} {\em {An introduction to fluid
  dynamics}\/}.  \publ{Cambridge university press}.

\bibitem[Betz(1920)]{Betz1920}
{\sc \au{Betz, A.}} \yr{1920}  \at{{Das maximum der theoretisch m{\"o}glichen
  ausn{\"u}tzung des windes durch windmotoren}}.  \jt{Zeitschrift f{\"u}r das
  gesamte Turbinenwesen}  \bvol{26}~(307-309),  \pg{8}.

\bibitem[Calaf {\em et~al.\/}(2010)Calaf, Meneveau \& Meyers]{Calaf2010}
{\sc \au{Calaf, M.}, \au{Meneveau, C.} \& \au{Meyers, J.}} \yr{2010}
  \at{{Large eddy simulation study of fully developed wind-turbine array
  boundary layers}}.  \jt{Physics of Fluids}  \bvol{22}~(1),  \pg{1--16}.

\bibitem[{CFD Module User's Guide}(2015)]{COMSOLCFD2016}
{\sc \au{{CFD Module User's Guide}}} \yr{2015} {\em {COMSOL Multiphysics® v.
  5.2.}\/}.  \publ{Stockholm, Sweden: COMSOL AB}.

\bibitem[Churchfield {\em et~al.\/}(2015)Churchfield, Fleming, Bulder \&
  White]{Churchfield2015}
{\sc \au{Churchfield, M.~J.}, \au{Fleming, P.}, \au{Bulder, B.} \& \au{White,
  S.~M.}} \yr{2015} {Wind Turbine Wake-Redirection Control at the Fishermen's
  Atlantic City Windfarm}.  \bt{In {\em {Offshore Technology Conference}\/}}.
  Offshore Technology Conference.

\bibitem[Corten {\em et~al.\/}(2007)Corten, Lindenburg \& Schaak]{Corten2007}
{\sc \au{Corten, G.~P.}, \au{Lindenburg, K.} \& \au{Schaak, P.}} \yr{2007}
  {Assembly of energy flow collectors, such as windpark, and method of
  operation}. US Patent 7,299,627.

\bibitem[Cummins(2013)]{Cummins2013}
{\sc \au{Cummins, P.~F.}} \yr{2013}  \at{{The extractable power from a split
  tidal channel: An equivalent circuit analysis}}.  \jt{Renewable Energy}
  \bvol{50},  \pg{395--401}.

\bibitem[Dabiri(2011)]{Dabiri2011a}
{\sc \au{Dabiri, J.~O.}} \yr{2011}  \at{{Potential order-of-magnitude
  enhancement of wind farm power density via counter-rotating vertical-axis
  wind turbine arrays}}.  \jt{Journal of Renewable and Sustainable Energy}
  \bvol{3}~(4),  \pg{1--12}.

\bibitem[Divett {\em et~al.\/}(2013)Divett, Vennell \& Stevens]{Divett2013}
{\sc \au{Divett, T.~A.}, \au{Vennell, R.} \& \au{Stevens, C.}} \yr{2013}
  \at{{Channel scale tuning of large tidal turbine arrays using large eddy
  simulations with adaptive mesh}}.  \jt{EWTEC 2013 Proceedings}  \bvol{86},
  \pg{1394--1405}.

\bibitem[Eaton {\em et~al.\/}(2014)Eaton, Bateman, Hauberg \&
  Wehbring]{Eaton2014}
{\sc \au{Eaton, J.~W.}, \au{Bateman, D.}, \au{Hauberg, S.} \& \au{Wehbring,
  R.}} \yr{2014} {\em {GNU Octave version 3.8.1 manual: a high-level
  interactive language for numerical computations}\/}.  \publ{CreateSpace
  Independent Publishing Platform}.

\bibitem[Fleming {\em et~al.\/}(2014)Fleming, Gebraad, Lee, van Wingerden,
  Johnson, Churchfield, Michalakes, Spalart \& Moriarty]{Fleming2014}
{\sc \au{Fleming, P.~A.}, \au{Gebraad, P. M.~O.}, \au{Lee, S.}, \au{van
  Wingerden, J-W.}, \au{Johnson, K.}, \au{Churchfield, M.}, \au{Michalakes,
  J.}, \au{Spalart, P.} \& \au{Moriarty, P.}} \yr{2014}  \at{{Evaluating
  techniques for redirecting turbine wakes using SOWFA}}.  \jt{Renewable
  Energy}  \bvol{70},  \pg{211--218}.

\bibitem[Frandsen(1992)]{Frandsen1992}
{\sc \au{Frandsen, S.}} \yr{1992}  \at{{On the wind speed reduction in the
  center of large clusters of wind turbines}}.  \jt{Journal of Wind Engineering
  and Industrial Aerodynamics}  \bvol{39}~(1-3),  \pg{251--265}.

\bibitem[Garrett \& Cummins(2005)]{Garrett2005}
{\sc \au{Garrett, C.} \& \au{Cummins, P.}} \yr{2005}  \at{{The power potential
  of tidal currents in channels.}}  \jt{Proc. R. Soc. A Math. Phys. Eng. Sci.}
  \bvol{461}~(2060),  \pg{2563--2572}.

\bibitem[Garrett \& Cummins(2007)]{Garrett2007}
{\sc \au{Garrett, C.} \& \au{Cummins, P.}} \yr{2007}  \at{{The efficiency of a
  turbine in a tidal channel}}.  \jt{Journal of Fluid Mechanics}  \bvol{588},
  \pg{243--251}.

\bibitem[Islam {\em et~al.\/}(2008)Islam, Ting \& Fartaj]{Islam2008}
{\sc \au{Islam, M.}, \au{Ting, D.~{S-K}} \& \au{Fartaj, A.}} \yr{2008}
  \at{{Aerodynamic models for Darrieus-type straight-bladed vertical axis wind
  turbines}}.  \jt{Renewable and Sustainable Energy Reviews}  \bvol{12}~(4),
  \pg{1087--1109}.

\bibitem[Jeffcoate {\em et~al.\/}(2016)Jeffcoate, Trevor \&
  Elsaesser]{Jeffcoate2016}
{\sc \au{Jeffcoate, P.}, \au{Trevor, W.} \& \au{Elsaesser, B.}} \yr{2016}
  \at{{Field tests of multiple 1/10 scale tidal turbines in steady flows}}.
  \jt{Renewable Energy}  \bvol{87}~(1),  \pg{240--252}.

\bibitem[{Jim{\'e}nez, {\'A}. and Crespo, A. and Migoya,
  E.}(2010)]{Jimenez2010}
{\sc \au{{Jim{\'e}nez, {\'A}. and Crespo, A. and Migoya, E.}}} \yr{2010}
  \at{{Application of a LES technique to characterize the wake deflection of a
  wind turbine in yaw}}.  \jt{Wind energy}  \bvol{13}~(6),  \pg{559--572}.

\bibitem[Joukowsky(1920)]{Joukowsky1920}
{\sc \au{Joukowsky, N.~E.}} \yr{1920}  \at{{Windmill of the NEJ type}}.
  \jt{Transactions of the Central Institute for Aero-hydrodynamics of Moscow}
  \bvol{1},  \pg{57}.

\bibitem[Kinzel {\em et~al.\/}(2012)Kinzel, Mulligan \& Dabiri]{Kinzel2012a}
{\sc \au{Kinzel, M.}, \au{Mulligan, Q.} \& \au{Dabiri, J.~O.}} \yr{2012}
  \at{{Energy exchange in an array of vertical-axis wind turbines}}.
  \jt{Journal of Turbulence}  \bvol{13}~(38),  \pg{N38}.

\bibitem[Kirshen {\em et~al.\/}(2016)Kirshen, Mandre \& Weiss]{Kirshen2016}
{\sc \au{Kirshen, L.}, \au{Mandre, S.} \& \au{Weiss, E.}} \yr{2016} {A model
  for tracking the wakes of vertical axis turbines}.
  doi.org/10.1103/APS.DFD.2016.GFM.P0043.

\bibitem[Launder \& Spalding(1974)]{Launder1974}
{\sc \au{Launder, Brian~Edward} \& \au{Spalding, Dudley~Brian}} \yr{1974}
  \at{The numerical computation of turbulent flows}.  \jt{Computer methods in
  applied mechanics and engineering}  \bvol{3}~(2),  \pg{269--289}.

\bibitem[Lighthill(1986)]{Lighthill1986}
{\sc \au{Lighthill, J.}} \yr{1986}  \at{{An informal introduction to
  theoretical fluid mechanics}} .

\bibitem[Lissaman(1979)]{Lissaman1979}
{\sc \au{Lissaman, P. B.~S.}} \yr{1979}  \at{{Energy Effectiveness of Arbitrary
  Arrays of Wind Turbines}}.  \jt{Journal of Energy}  \bvol{3}~(6),
  \pg{323--328}.

\bibitem[MacKay(2008)]{Mackay2008}
{\sc \au{MacKay, D.}} \yr{2008} {\em {Sustainable Energy-without the hot
  air}\/}.  \publ{UIT Cambridge}.

\bibitem[Mandre {\em et~al.\/}(2016)Mandre, Mangan, Derecktor \&
  Winckler]{Mandre2016a}
{\sc \au{Mandre, S.}, \au{Mangan, N.~M.}, \au{Derecktor, T.} \& \au{Winckler,
  S.}} \yr{2016} {A comparison of hydrokinetic turbines forming a vertical
  fence along the length of a river or tidal channel with a conventional
  rectangular turbine array}.  \bt{In {\em {Proceedings of the 4th Marine
  Energy Technology Symposium}\/}}.

\bibitem[Mangan \& Mandre(2016)]{Mangan2016a}
{\sc \au{Mangan, N.~M.} \& \au{Mandre, S.}} \yr{2016} {Optimal distribution of
  riverine turbines in a linear array with systematic flow manipulation}.
  \bt{In {\em {Proceedings of the 24$^\text{th}$ International Congress of
  Theoretical and Applied Mechanics}\/}}.

\bibitem[Meyers \& Meneveau(2013)]{Meyers2013}
{\sc \au{Meyers, J.} \& \au{Meneveau, C.}} \yr{2013}  \at{{Flow visualization
  using momentum and energy transport tubes and applications to turbulent flow
  in wind farms}}.  \jt{Journal of Fluid Mechanics}  \bvol{715},
  \pg{335--358}.

\bibitem[Newman(1977)]{Newman1977}
{\sc \au{Newman, B.~G.}} \yr{1977}  \at{{The spacing of wind turbines in large
  arrays}}.  \jt{Energy Conversion}  \bvol{16}~(4),  \pg{169--171}.

\bibitem[Newman(1986)]{Newman1986}
{\sc \au{Newman, B.~G.}} \yr{1986}  \at{{Multiple actuator-disc theory for wind
  turbines}}.  \jt{Journal of Wind Engineering and Industrial Aerodynamics}
  \bvol{24}~(3),  \pg{215--225}.

\bibitem[Okulov \& van Kuik(2012)]{Okulov2012}
{\sc \au{Okulov, V.~L.} \& \au{van Kuik, G. A.~M.}} \yr{2012}  \at{{The
  Betz-Joukowsky limit: on the contribution to rotor aerodynamics by the
  British, German and Russian scientific schools}}.  \jt{Wind Energy}
  \bvol{15}~(2),  \pg{335--344}.

\bibitem[Pedlosky(2013)]{Pedlosky2013}
{\sc \au{Pedlosky, Joseph}} \yr{2013} {\em {Geophysical fluid dynamics}\/}.
  \publ{Springer Science \& Business Media}.

\bibitem[Port{\'e}-Agel {\em et~al.\/}(2013)Port{\'e}-Agel, Wu \&
  Chen]{Porte-Agel2013}
{\sc \au{Port{\'e}-Agel, F.}, \au{Wu, Y-T.} \& \au{Chen, C-H.}} \yr{2013}
  \at{{A Numerical Study of the Effects of Wind Direction on Turbine Wakes and
  Power Losses in a Large Wind Farm}}.  \jt{Energies}  \bvol{6}~(10),
  \pg{5297--5313}.

\bibitem[Saffman(1992)]{Saffman1992}
{\sc \au{Saffman, P.~G.}} \yr{1992} {\em {Vortex dynamics}\/}.  \publ{Cambridge
  university press}.

\bibitem[Thresher {\em et~al.\/}(2007)Thresher, Robinson \&
  Veers]{Thresher2007}
{\sc \au{Thresher, R.}, \au{Robinson, M.} \& \au{Veers, P.}} \yr{2007}  \at{{To
  capture the wind}}.  \jt{IEEE Power and Energy Magazine}  \bvol{5}~(6),
  \pg{34--46}.

\bibitem[Vennell(2010)]{Vennell2010}
{\sc \au{Vennell, R.}} \yr{2010}  \at{{Tuning turbines in a tidal channel}}.
  \jt{Journal of Fluid Mechanics}  \bvol{663},  \pg{253--267}.

\bibitem[Vennell(2011)]{Vennell2011}
{\sc \au{Vennell, R.}} \yr{2011}  \at{{Estimating the power potential of tidal
  currents and the impact of power extraction on flow speeds}}.  \jt{Renewable
  Energy}  \bvol{36}~(12),  \pg{3558--3565}.

\bibitem[Vennell {\em et~al.\/}(2015)Vennell, Funke, Draper, Stevens \&
  Divett]{Vennell2015}
{\sc \au{Vennell, R.}, \au{Funke, S.~W.}, \au{Draper, S.}, \au{Stevens, C.} \&
  \au{Divett, T.}} \yr{2015}  \at{{Designing large arrays of tidal turbines: A
  synthesis and review}}.  \jt{Renewable and Sustainable Energy Reviews}
  \bvol{41},  \pg{454--472}.

\bibitem[VerHulst \& Meneveau(2015)]{VerHulst2015}
{\sc \au{VerHulst, C.} \& \au{Meneveau, C.}} \yr{2015}  \at{{Altering Kinetic
  Energy Entrainment in Large Eddy Simulations of Large Wind Farms Using
  Unconventional Wind Turbine Actuator Forcing}}.  \jt{Energies}  \bvol{8}~(1),
   \pg{370--386}.

\bibitem[Wagenaar {\em et~al.\/}(2012)Wagenaar, Machielse \&
  Schepers]{Wagenaar2012}
{\sc \au{Wagenaar, J.~W.}, \au{Machielse, L. A.~H.} \& \au{Schepers, J.~G.}}
  \yr{2012}  \bt{{Controlling Wind in ECN's Scaled Wind Farm}}.  \pg{pp. 3--4}.

\bibitem[Whittlesey {\em et~al.\/}(2010)Whittlesey, Liska \&
  Dabiri]{Whittlesey2010}
{\sc \au{Whittlesey, R.~W.}, \au{Liska, S.} \& \au{Dabiri, J.~O.}} \yr{2010}
  \at{{Fish schooling as a basis for vertical axis wind turbine farm design}}.
  \jt{Bioinspiration \& biomimetics}  \bvol{5}~(3),  \pg{035005}.

\end{thebibliography}

\end{document}